\documentstyle{mn}

\newif\ifAMStwofonts

\ifoldfss
  \ifCUPmtlplainloaded \else
    \NewTextAlphabet{textbfit} {cmbxti10} {}
    \NewTextAlphabet{textbfss} {cmssbx10} {}
    \NewMathAlphabet{mathbfit} {cmbxti10} {} 
    \NewMathAlphabet{mathbfss} {cmssbx10} {} 
  \fi
  \ifAMStwofonts
    \ifCUPmtlplainloaded \else
      \NewSymbolFont{upmath} {eurm10}
      \NewSymbolFont{AMSa} {msam10}
      \NewMathSymbol{\upi}     {0}{upmath}{19}
      \NewMathSymbol{\umu}     {0}{upmath}{16}
      \NewMathSymbol{\upartial}{0}{upmath}{40}
      \NewMathSymbol{\leqslant}{3}{AMSa}{36}
      \NewMathSymbol{\geqslant}{3}{AMSa}{3E}

      \let\leq=\leqslant \let\le=\leqslant
      \let\geq=\geqslant \let\ge=\geqslant
    \fi
  \fi
\fi 

\ifnfssone
  \newmathalphabet{\mathit}
  \addtoversion{normal}{\mathit}{cmr}{m}{it}
  \addtoversion{bold}{\mathit}{cmr}{bx}{it}
  \newmathalphabet{\mathbfit} 
  \addtoversion{normal}{\mathbfit}{cmr}{bx}{it}
  \addtoversion{bold}{\mathbfit}{cmr}{bx}{it}
  \newmathalphabet{\mathbfss} 
  \addtoversion{normal}{\mathbfss}{cmss}{bx}{n}
  \addtoversion{bold}{\mathbfss}{cmss}{bx}{n}
  \ifAMStwofonts
    \ifCUPmtlplainloaded \else
      \UseAMStwoboldmath
      \makeatletter
      \new@mathgroup\upmath@group
      \define@mathgroup\mv@normal\upmath@group{eur}{m}{n}
      \define@mathgroup\mv@bold\upmath@group{eur}{b}{n}
      \edef\UPM{\hexnumber\upmath@group}
      \new@mathgroup\amsa@group
      \define@mathgroup\mv@normal\amsa@group{msa}{m}{n}
      \define@mathgroup\mv@bold\amsa@group{msa}{m}{n}
      \edef\AMSa{\hexnumber\amsa@group}
      \makeatother
      \mathchardef\upi="0\UPM19
      \mathchardef\umu="0\UPM16
      \mathchardef\upartial="0\UPM40
      \mathchardef\leqslant="3\AMSa36
      \mathchardef\geqslant="3\AMSa3E

      \let\leq=\leqslant \let\le=\leqslant
      \let\geq=\geqslant \let\ge=\geqslant
    \fi
  \fi
\fi 

\ifnfsstwo
  \DeclareMathAlphabet{\mathbfit}{OT1}{cmr}{bx}{it}
  \SetMathAlphabet\mathbfit{bold}{OT1}{cmr}{bx}{it}
  \DeclareMathAlphabet{\mathbfss}{OT1}{cmss}{bx}{n}
  \SetMathAlphabet\mathbfss{bold}{OT1}{cmss}{bx}{n}
  \ifAMStwofonts
    \ifCUPmtlplainloaded \else
      \DeclareSymbolFont{UPM}{U}{eur}{m}{n}
      \SetSymbolFont{UPM}{bold}{U}{eur}{b}{n}
      \DeclareSymbolFont{AMSa}{U}{msa}{m}{n}
      \DeclareMathSymbol{\upi}{0}{UPM}{"19}
      \DeclareMathSymbol{\umu}{0}{UPM}{"16}
      \DeclareMathSymbol{\upartial}{0}{UPM}{"40}
      \DeclareMathSymbol{\leqslant}{3}{AMSa}{"36}
      \DeclareMathSymbol{\geqslant}{3}{AMSa}{"3E}

      \let\leq=\leqslant \let\le=\leqslant
      \let\geq=\geqslant \let\ge=\geqslant
    \fi
  \fi
\fi 

\ifCUPmtlplainloaded \else
  \ifAMStwofonts \else 
    \def\upi{\pi}
    \def\umu{\mu}
    \def\upartial{\partial}
  \fi
\fi

\title[Radio sources in the 2dFGRS] {Radio sources in the 2dF Galaxy Redshift 
Survey -- II. Local radio luminosity functions for AGN and star--forming 
galaxies at 1.4\,GHz}

\author[Sadler et al. ]{
\parbox[t]{\textwidth}{
Elaine M.\ Sadler$^1$, 
Carole A.\ Jackson$^2$, 
Russell D.\ Cannon$^3$,
Vincent J.\ McIntyre$^4$, 
Tara Murphy$^1$, 
Joss Bland--Hawthorn$^3$,
Terry Bridges$^3$, 
Shaun Cole$^5$, 
Matthew Colless$^2$, 
Chris Collins$^6$, 
Warrick Couch$^7$, 
Gavin Dalton$^8$,
Roberto De Propris$^7$,
Simon P.\ Driver$^9$, 
George Efstathiou$^{10}$, 
Richard S.\ Ellis$^{11}$, 
Carlos S.\ Frenk$^5$, 
Karl Glazebrook$^{12}$, 
Ofer Lahav$^{10}$, 
Ian Lewis$^3$, 
Stuart Lumsden$^{13}$, 
Steve Maddox$^{14}$,
Darren Madgwick$^{10}$,
Peder Norberg$^5$,
John A.\ Peacock$^{15}$, 
Bruce A.\ Peterson$^2$, 
Will Sutherland$^8$,
Keith Taylor$^{11}$}
\vspace*{6pt} \\ 
$^1$School of Physics, University of Sydney, NSW 2006, Australia\\
$^2$Research School of Astronomy \& Astrophysics, The Australian 
    National University, Weston Creek, ACT 2611, Australia \\
$^3 $Anglo-Australian Observatory, P.O.\ Box 296, Epping, NSW 2121,
    Australia\\  
$^4$Australia Telescope National Facility, CSIRO, P.O. Box 76, Epping, 
    NSW 2121, Australia \\
$^5$Department of Physics, South Road, Durham DH1 3LE, UK \\
$^6$Astrophysics Research Institute, Liverpool John Moores University,  
    Twelve Quays House, Birkenhead, L14 1LD, UK \\
$^7$Department of Astrophysics, University of New South Wales, Sydney, 
    NSW 2052, Australia \\
$^8$Department of Physics, Keble Road, Oxford OX1 3RH, UK \\
$^9$School of Physics and Astronomy, North Haugh, St Andrews, Fife, 
    KY6 9SS, UK \\
$^{10}$Institute of Astronomy, University of Cambridge, Madingley Road,
    Cambridge CB3 0HA, UK \\
$^{11}$Department of Astronomy, Caltech, Pasadena, CA 91125, USA \\
$^{12}$Department of Physics \& Astronomy, Johns Hopkins University,
       Baltimore, MD 21218-2686, USA \\
$^{13}$Department of Physics, University of Leeds, Woodhouse Lane,
       Leeds, LS2 9JT, UK \\
$^{14}$School of Physics \& Astronomy, University of Nottingham,
       Nottingham NG7 2RD, UK \\
$^{15}$Institute for Astronomy, University of Edinburgh, Royal Observatory, 
       Blackford Hill, Edinburgh EH9 3HJ, UK \\
}
 
\pagerange{\pageref{firstpage}--\pageref{lastpage}}
\pubyear{2002}

\begin{document}

\maketitle

\label{firstpage}

\begin{abstract}
We have cross--matched the 1.4\,GHz NRAO VLA Sky Survey (NVSS) with the 
first 210 fields observed in the 2dF Galaxy Redshift Survey (2dFGRS), 
covering an effective area of 325 square degrees (about 20\% of the final 
2dFGRS area).  This yields a set of optical spectra of 912 candidate NVSS 
counterparts, of which we identify 757 as genuine radio IDs --- the 
largest and most homogeneous set of radio--source spectra ever obtained. 
The 2dFGRS radio sources span the redshift range $z=0.005$ to 0.438, 
and are a mixture of active galaxies (60\%) and star--forming galaxies 
(40\%). About 25\% of the 2dFGRS radio sources are spatially resolved 
by NVSS, and the sample includes three giant radio galaxies with projected 
linear size greater than 1\,Mpc. 
The high quality of the 2dF spectra means we can usually distinguish 
unambiguously between AGN and star--forming galaxies.  
We have made a new determination of the local radio luminosity function at
1.4\,GHz for both active and star--forming galaxies, and derive a local
star--formation density of $0.022\pm0.004$ M$_\odot$ yr$^{-1}$ Mpc$^{-3}$ 
(H$_0$=50\,km\,s$^{-1}$\,Mpc$^{-1}$).  
\end{abstract}


\vspace*{1cm}

\section{Introduction}
Radio source surveys are ideal tools for studying the distant universe, 
since they are unaffected by dust obscuration and detect large numbers of 
galaxies over a wide span of cosmic epochs (the median redshift of galaxies 
detected in radio surveys is typically $z\simeq1$; Condon 1989). 

At 1.4\,GHz flux densities above about 50\,mJy, more than 95\% of radio
sources are classical radio galaxies and quasars powered by active nuclei 
(AGN).  Below 50\,mJy, the AGN proportion declines and an increasing fraction
of the faint radio source population is identified with star--forming galaxies
(e.g. Condon 1989, 1992). These are usually disk galaxies, sometimes
interacting with neighbours, in which the radio emission arises mainly through
synchrotron emission from relativistic electrons accelerated by supernova
explosions. Thus radio surveys to levels of a few mJy probe both the 
AGN population and a population of star--forming galaxies.   
It is important to be able to separate these, in order to determine 
the local space density and redshift evolution of each population.  

A new generation of radio imaging surveys (NVSS, Condon et al.\ 1998; 
FIRST, Becker, White \& Helfand 1995; WENSS, Rengelink et al.\ 1997; 
SUMSS, Bock, Large \& Sadler 1999) 
is now covering the whole sky to sensitivities of a few mJy. 
Radio source counts from such surveys potentially yield important 
information on the cosmological evolution of active and star--forming 
galaxies (e.g. Longair 1966, Jauncey 1975, Wall et al.\ 1980), but their
interpretation is strongly model--dependent.  The scientific return from 
large radio surveys is enormously increased if the optical counterparts 
of the radio sources can be identified, their optical spectra classified 
(as AGN, starburst galaxy, etc.) and their redshift distribution measured.  
In the past, however, this was a slow and tedious process which could only 
be carried out for relatively small samples. 

Now, fibre--fed optical spectrographs make it possible to carry out
spectroscopy of complete samples of hundreds of thousands of galaxies 
in the local universe. 
The Anglo--Australian Observatory's Two--degree Field (2dF) spectrograph can 
observe up to 400 galaxies simultaneously over a 2$^\circ$-diameter
region of sky (Lewis et al.\ 2001, see also {\tt www.aao.gov.au/2df/}).  
A Six--degree Field (6dF) spectrograph will be
commissioned on the AAO's Schmidt Telescope in 2001, with 150 fibres 
over a 6$^\circ$-diameter field, and in the northern hemisphere the Sloan 
Digital Sky Survey (SDSS; York et al.\ 2000) has begun a program of 
spectroscopy of 10$^6$ galaxies.  Cross--matching radio continuum surveys 
with these new optical redshift surveys will provide redshifts and 
spectroscopic data for tens of thousands of local radio--emitting 
galaxies, rather than the few hundred available at present. 

Two recent studies show the potential of these new redshift surveys.  
Machalski \& Condon (1999) identified 1157 galaxies in the Las Campanas 
Redshift Survey (LCRS) with NVSS radio sources above 2.5\,mJy at 
1.4\,GHz.  They attempted to determine the radio and infrared properties 
of galaxies in the LCRS redshift range of $z=0.05 - 0.2$, but had
difficulties because the LCRS was sparsely sampled, with optical
spectra only being taken for about one galaxy in three.  Nevertheless, 
Machalski \& Godlowski (2000) used the LCRS to derive the local radio 
luminosity function for both AGN and star--forming galaxies, and to test 
for evolution over the LCRS redshift range.  

Sadler et al.\ (1999) cross--matched the NVSS radio catalogue with the first
thirty fields observed in the 2dFGRS, and found that it was
usually straightforward to tell from the optical spectra whether the radio
emission arose from star formation or an AGN (unlike LCRS, the 2dFGRS has a
spectroscopic completeness of 95\%). 

The present paper is the second in a series analysing the properties 
of NVSS radio sources which are identified with galaxies in the 2dF 
Galaxy Redshift Survey (2dFGRS; Colless 1999, Colless et al.\ 2001, 
see also {\tt www.mso.edu.au/2dFGRS/}).  When complete, the 2dFGRS 
will yield good--quality optical spectra for up to 4000 radio--emitting 
galaxies in the redshift range $z=0$ to 0.4.  Such a large sample should 
allow us to disentangle the effects of age, orientation and luminosity 
in the local AGN population, as well as providing a definitive measurement 
of the local radio luminosity function (RLF) for active and star--forming 
galaxies.  Our first paper (Sadler et al.\ 1999; hereafter Paper I), 
was a preliminary investigation of a small sample of the early 2dFGRS data, 
and showed that the 2dFGRS radio--source population is composed of roughly 
similar numbers of AGN and star--forming galaxies.  
In this paper we analyse a 2dFGRS data set  which is almost an order of 
magnitude larger than in Paper I, and use it to derive the local radio 
luminosity function for active and star--forming galaxies at 1.4\,GHz.  

The spectra analysed in this paper are included in the 
first public release of 2dFGRS data, which took place in June 2001. 

Throughout this paper, we use $H_0$ = 50 km s$^{-1}$ 
Mpc$^{-1}$ and $\Omega_0$=1.

\section{Sample selection}
In this paper, we analyse 2dFGRS spectra from the first two years of the
survey, i.e. up to May 1999.  Observations made before October 1997 were
excluded because the instrument was still in a commissioning phase and the 
data are of variable quality.  The data set analysed here comprises 210 
fields, or about 20\% of the total 2dFGRS area.  The original selection of 
2dFGRS optical targets included all non-stellar objects brighter than 
$b_{\rm J}\simeq19.5$ from the photographic UKST Southern Sky Survey, and 
should involve no explicit biases so far as radio properties are concerned.

\subsection{Area covered} 
The 2dFGRS uses a tiling algorithm with variable overlap depending on the 
underlying galaxy density.  Some 2dFGRS fields are also shared with 
a parallel QSO redshift survey, so calculating the exact area of sky observed 
is not straightforward until the whole survey is complete. 

We therefore use the method described by Folkes et al.\ (1999) and estimate 
the area of sky covered by dividing the number of galaxies observed by the 
mean surface density of galaxies in the target list.  The data set observed 
in the period October 1997---May 1999 inclusive contains 58,454 2dFGRS targets. 
Taking the Folkes et al.\ surface density of 180 galaxies deg$^{-2}$ to the
survey limit of $b_{\rm J} = 19.45$\,mag.\ gives an effective area of 325 
deg$^2$ for the data set we will analyse here. 

\subsection{Radio source identifications} 
We cross--matched the 2dFGRS catalogue positions with the 
NVSS radio catalogue and took all matches with position offsets of 
15 arcsec or less as candidate radio detections.  This yielded a total 
of 927 observations of 903 targets. Not all these will be true identifications 
--- as discussed in Paper I, we expect most unresolved radio sources 
with offsets of 10 arcsec or less to be true associations with 2dFGRS 
galaxies, along with a smaller (and quantifiable) fraction of sources with 
offsets of 10--15\,arcsec.

There are also three classes of objects for which the situation is more
complex: (i) extended (resolved) NVSS radio sources, (ii) radio sources
which are resolved by NVSS into two or more distinct components, 
and (iii) radio sources associated with nearby bright
galaxies (b$_{\rm J} < 17$\,mag).  As we discuss in Section 4, these 
received extra attention to determine whether to accept a candidate ID 
as a correct one. 

For completeness, and because others may wish to use our data with
different selection criteria for radio--source IDs, we have tabulated 2dFGRS
spectral types and redshifts for all 912 candidate radio--source IDs with
radio--optical offsets of 15.0\,arcsec or less. We showed in Paper I that, 
with the exception of a small number of double NVSS radio sources (see 
Section 4.2), there should be very few correct IDs with radio--optical 
offsets larger than 15\,arcsec.  This is confirmed in Fig. 1, which shows 
that by 15\,arcsec from the 2dF position the number of candidate detections 
falls to the level expected from chance coincidences alone. 

\begin{figure}

\vspace*{10cm}
\includegraphics{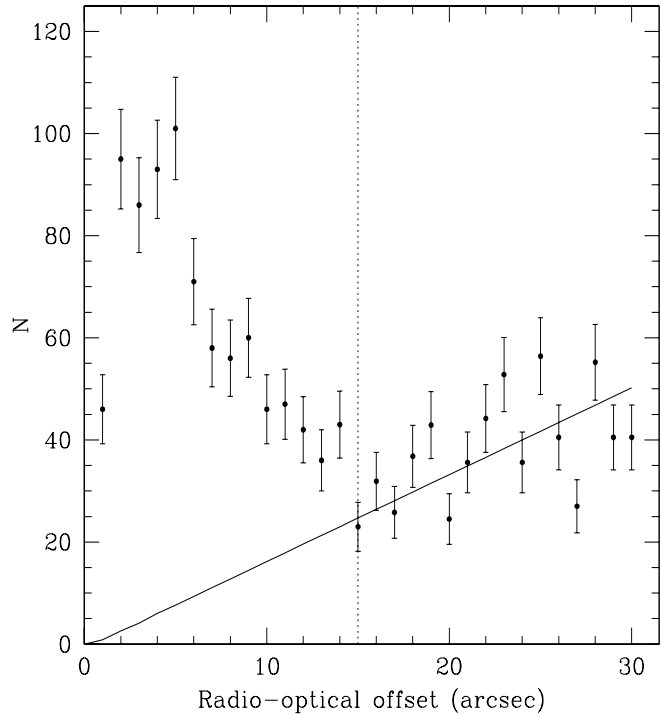}

\caption{Number of candidate NVSS detections of 2dFGRS galaxies plotted 
against the offset $\Delta$ between radio and optical positions.  
The solid line shows the number of matches expected by chance from a 
sample of 58,454 2dFGRS galaxies, assuming that NVSS radio sources are 
randomly distributed on the sky with a surface density of 60\,deg$^{-2}$. 
Note that at a separation of 15\,arcsec or more, the number of matches 
declines to that which would be expected by chance, so we adopted a 
separation cutoff of 15\,arcsec for candidate 2dFGRS radio sources.  }
\end{figure} 

We also note that the 2dFGRS input catalogue is incomplete for galaxies 
brighter than $b_{\rm J}\simeq14$, and so some bright, nearby galaxies 
are missing from the 2dFGRS.  Almost all 
these missing galaxies should already have redshift measurements, and 
can be added back into the sample once the entire 2dFGRS is complete. 
In calculating luminosity functions (see Section 8), we correct for the 
absence of these bright galaxies.

\section{Optical spectra}
As noted earlier, the spectra analysed in this paper were collected 
as the 2dFGRS progressed, and cover the period to May 1999.  Most of 
these spectra are included in the first (mid--2001) public release of 
2dFGRS data (see Colless et al.\ 2001 for details of the 
public--release database). In a few cases, a new spectrum taken after 
May 1999 may be better than the one used here and will replace it in the 
final database. 

Although in most cases the public--release data 
are identical to those analysed here, some values 
of $b_{\rm J}$, $z$ and Q listed in Table 1 may change. 
This is because there have been revisions of the $b_{\rm J}$\ magnitude 
scale, and a re-reduction of all the early 2dFGRS redshifts, since 
we began our analysis.   In the great majority of cases, the database 
revisions produced no significant difference, especially for spectra 
with Q$\geq$3. Small changes in the values of $b_{\rm J}$ or $z$ for 
individual galaxies will not affect the overall results of the analysis 
which follows. 

\subsection{Spectral classification}
2dFGRS spectra of each of the 912 candidate radio--source IDs 
were examined and classified visually by one of us (EMS). 
As in Paper I, each spectrum was classified as either `AGN' or 
`star--forming' (SF) based on its 2dF spectrum. AGN galaxies have 
either an absorption--line spectrum like that of a giant elliptical 
galaxy (these we classed `Aa'); an absorption--line spectrum 
with weak LINER--like emission lines (class `Aae'), or a
stellar continuum dominated by nebular emission lines such as [OII] 
and [OIII] which are strong compared with any Balmer--line emission 
(class `Ae').  
`SF' galaxies are those where strong, narrow emission lines of H$\alpha$ 
and (usually) H$\beta$ dominate the spectrum.   Objects where the spectra 
were too noisy for reliable classification are classed as `??'. 

There are several independent tests of the reliability of our visual 
classification of the spectra: (a) the visual clasifications generally 
agree well with the results of Principal Components Analysis (PCA) 
methods, as we will show in Section 3.3, (b) for galaxies with optical 
emission lines, our Ae/SF classifications agree well with those derived 
by Jackson \& Londish (2000) from diagnostic emission--line ratios, 
and (c) most galaxies which we classify as SF are detected by IRAS and 
fall on the radio---FIR correlation (see section 6), whereas none of 
the galaxies we classified as Aa are detected by IRAS.   
Thus we are confident that our visual classification of the 2dFGRS 
spectra allows us to distinguish AGN and star--forming galaxies in a 
consistent and reliable way. 

\subsection{Data table}
Table 1 lists the entire sample of 912 candidate radio IDs, i.e. the 903 
2dFGRS targets whose catalogued position is within 15\,arcsec of the catalogued 
position of an NVSS radio source, together with a further nine extended and 
double radio sources which were identified separately as described in 
Section 4.2. 
The table columns are as follows: 
\begin{itemize}
\item[(1)]
2dFGRS galaxy name (this is the name used in the 2dFGRS database). 
An asterisk following the name indicates that this galaxy appears 
in the Notes at the end of Table 1. 
\item[(2)]
Other name --- we have cross--identified with other galaxy catalogues where 
possible. 
\item[(3)] 
The optical position (J2000.0) at which the 2dFGRS fibre was placed. 
\item[(4)]
The offset between 2dFGRS (optical) and NVSS (radio) 
positions, in arcsec. 
\item[(5)]
Total blue ($b_{\rm J}$) magnitude, from the 2dFGRS database. 
The magnitudes listed here are taken from the May 2000 version of 
the database. 
\item[(6)]
Total radio flux density at 1.4\,GHz, from the NVSS catalogue (Condon 
et al.\ 1998). Where a source is split into more than one component in the 
NVSS catalogue, this is indicated in the Notes at the end of the table  
and the value quoted here is the sum of all the components. 
\item[(7)]
IRAS 60\,$\mu$m flux density (in Jy), where listed in the NASA Extragalactic 
Database (NED).  Although IRAS surveyed 97\% of the sky, the incompleteness 
near the Galactic poles is larger than in other areas.  Based on the 
IRAS sky coverage plots given by Beichman et al.\ (1988), about 10\% of 
the 2dFGRS survey area had either one or no IRAS scans and so was not 
catalogued.  The main 2dFGRS areas missed by IRAS lie in the region between 
10 and 11\,h RA in the northern zone and 23 and 0\,h RA in the southern zone, 
but there are also smaller gaps elsewhere.  
\item[(8)]
Heliocentric redshift, from 2dFGRS observations unless otherwise indicated 
in the Notes. 
\item[(9)]
2dFGRS spectrum quality code Q, where Q=4 or 5 are excellent--quality 
spectra, Q=3 acceptable and Q=0, 1 or 2 poor--quality.  Galaxies with 
Q$<$3 are excluded from further analysis because their redshifts are 
highly uncertain. Checks against repeat observations and published 
redshifts show that 2dFGRS redshifts with Q=3 are about 90\% reliable 
and those with Q=4 or 5 are 99\% reliable (Colless et al.\ 2001). 
\item[(10)]
Spectral class, based on our visual classification. \end{itemize}

\subsection{Comparison of visual and PCA spectral classifications  } 
We would eventually like to compare the 2dFGRS radio--emitting galaxies
with the parent sample from which they are drawn (in order to answer
questions like ``what fraction of Seyfert 1 and 2 galaxies are
radio--loud, and how does this vary with redshift?''), so we compared 
our simple visual classification of the spectra (as Aa, Ae or SF) with
the Principal Components Analysis (PCA) methods developed for use with 
the 2dFGRS by Folkes et al.\ (1999) and Madgwick et al.\ (2001).  
PCA is an automatic method
of classifying spectra which has the great advantages of being
objective, quantitative and easily applied to very large samples.  
A series of Principal Components or eigenspectra (PC1, PC2, ...) are
determined from the distribution of all the spectra in a very large
multi--dimensional space; most of the information in the spectra is
included in the first few PCs.  This means that each spectrum can be
well--represented by a set of just 3 or 4 numbers, corresponding to the
relative power in the dominant PCs. 

We find good agreement between the visual and automatic methods of 
classifying galaxy spectra. For example, in a plot of PC3 against PC1 
(Fig. 2) there is a clear separation between the different visual 
spectral types.  By combining similar plots involving PC2, it looks as 
if it will be possible to define an almost unique mapping between the 
two classification methods, making it easy to compare the properties of
the 1.5\% of the galaxies which are radio sources with the full 2dFGRS 
sample of up to 250,000 optical galaxies.

\begin{figure}

\vspace*{9cm}
\includegraphics{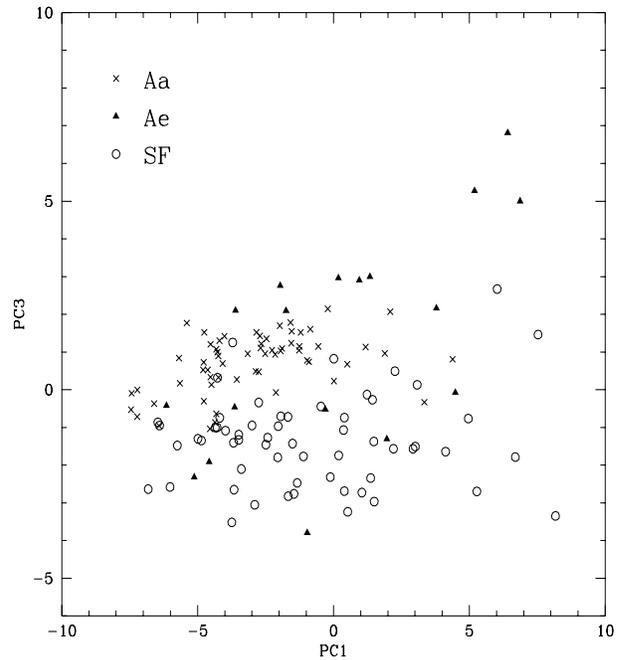}

\caption{Plots of the first and third PCA components from 2dF spectra of 
radio--emitting galaxies.  The three spectral classes from our visual 
classification are reasonably well--separated in this plot, though some 
Ae galaxies overlap the region populated by SF galaxies.  
The PCA data used in this preliminary analysis were taken from
an early version of the database and there have been some significant
revisions of the PCA scheme since then (Madgwick et al.\ 2001). }
\end{figure} 

\subsection{Spectral class and redshift} 
Fig. 3 shows how the composition of the 2dFGRS radio sources changes 
with redshift.  As we showed in Paper I, star--forming galaxies dominate 
the population at low redshift while active galaxies dominate above 
$z\simeq0.1$.  
This is because star--forming galaxies are low--luminosity radio sources 
(P$_{1.4} < 10^{23}$ W Hz$^{-1}$) which drop out of the sample when they 
fall below the NVSS radio detection limit. 
Conversely, most AGN have higher radio luminosities and remain in the sample
out to $z\simeq0.2-0.3$ where they finally drop below the 2dFGRS optical 
magnitude limit.  

\begin{figure}

\vspace*{9cm}
\includegraphics{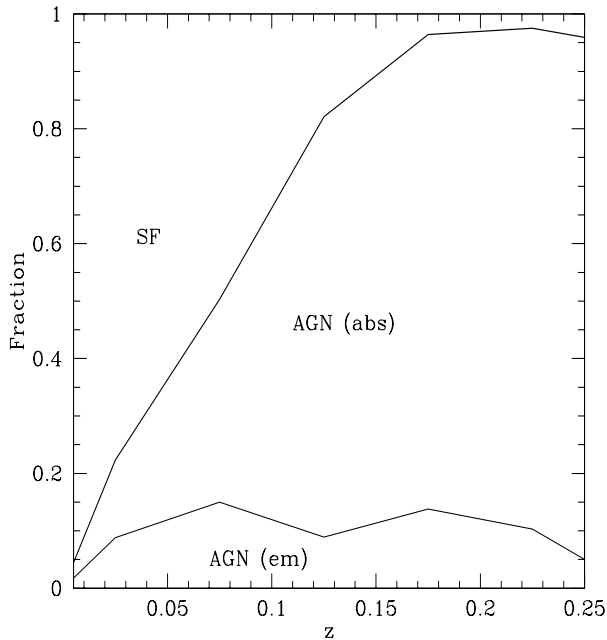}

\caption{Change in spectral mix with redshift for 2dFGRS/NVSS galaxies.  
SF indicates star--forming galaxies with HII-region optical spectra and AGN 
(em) and AGN (abs) indicate active galaxies with and without strong optical emission lines. } 
\end{figure} 

Interestingly, the fraction of emission-line AGN (Ae) galaxies remains roughly 
constant (at 10--15\%) throughout the sample volume even though the overall AGN 
fraction increases dramatically with redshift.  We need to keep in mind, 
however, that the 2\,arcsec diameter 2dF fibres include an increasing fraction 
of the total galaxy light for more distant galaxies (see Fig. 4). 
As a result, galaxies with emission--line nuclei will be easier to recognise 
at lower redshift, where there is less dilution from  the surrounding 
stellar galaxy, than at higher redshift.  We therefore assume that the 
probability of recognizing a galaxy as Ae rather than Aa varies with 
redshift, and so we combine the Aa and Ae classes in most of our later 
analysis. 

In star--forming galaxies, the line emission is expected to come 
mainly from an extended disk and dilution of the emission--line flux with 
redshift is less likely.  Indeed we might expect star--forming galaxies 
to be easier to recognise at higher redshift since the 2dF aperture includes 
a larger fraction of the total disk light.  

\begin{figure}

\vspace*{8cm}
\includegraphics{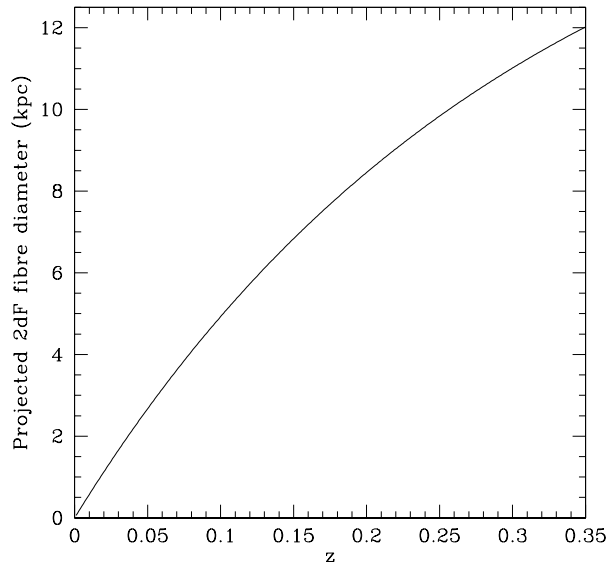}

\caption{Projected diameter of the 2dF fibres (in kpc) for galaxies over the 
redshift range $z = 0$ to 0.35. }
\end{figure}

\section{Radio--optical identifications } 
We now need to determine which of the candidate radio source identifications 
in Table 1 are real associations between a 2dFGRS galaxy and an NVSS
radio source.  This is not always straightforward --- as noted in paper I, 
we expect some chance coincidences even at the smallest radio--optical
separations. In Paper I, we used a simple 10\,arcsec cutoff as the criterion 
for association. 
Although this will include a few objects which are not real associations and
exclude some which are, the {\sl number}\ of such objects can be quantified
statistically and taken into account when calculating the luminosity function. 

We use a similar 10\,arcsec cutoff here, but also recognise that a simple 
cutoff may exclude some extended sources with complex radio structure. 
Thus we also inspected radio--optical overlays of all the extended sources in
the candidate list and accepted some of these as genuine IDs even though 
the radio--optical offset was larger than 10\, arcsec.  

\subsection{Extended radio sources} 
About 25\% of the candidate NVSS/2dFGRS radio sources are resolved by 
the 45\,arcsec NVSS beam, giving us an estimate of their linear size.  
In most such cases we still see a single radio source, but with an apparent 
size larger than 45\,arcsec in at least one dimension.  
For these galaxies, the NVSS radio position is calculated by fitting a 
two--dimensional Gaussian to the radio image.  If the fit is not perfect, 
the quoted position may have a much larger uncertainly than that of an 
unresolved radio source of similar flux density.  

For all extended NVSS sources with radio--optical offsets in the range 
10.0--15.0\,arcsec, we inspected overlays of the radio contours onto 
DSS images (see Fig. 5 for examples).  If the optical galaxy lay along 
the major axis of the extended radio source, or the radio emission appeared 
to be symmetric about the optical galaxy, we accepted this as a correct ID. 

\subsection{Double radio sources} 
A few NVSS radio sources are resolved into two or more distinct components, 
so we need to be able to recognise these, and to search for an optical ID 
near the radio centroid rather than at the position of each individual  component. 

We first identified all the candidate double radio sources in the NVSS
catalogue by using the semi--empirical link--length measure defined by
Magliocchetti et al.\ (1998). A pair of radio sources are defined to be
associated if: \\ 
(i) the ratio of their flux densities S$_1$ and S$_2$ lies 
in the range 0.25 $<$ (S$_1$/S$_2$) $<$ 4.0, {\it and}  \\
(ii) the projected separation of the two radio components is smaller 
than their link length $r$, which is calculated from the total flux density 
in mJy and defined as: 
\begin{equation}
r = [({\rm S}_1+{\rm S}_2)/100\,{\rm mJy}]^{0.5}\ \times 100\,
{\rm arcsec}. 
\end{equation}

This yielded a total of 525 candidate NVSS doubles in the 1700\,deg$^2$ 
region of sky covered by the 2dFGRS.  Of these, 59 had a 2dFGRS galaxy 
within 30\,arcsec of the radio centroid and 17 of these galaxies had spectra 
taken up to 1999 May.  These objects were then inspected to see whether the
optical galaxy lay on or near the radio axis of the NVSS double.  As a result,
eight new 2dFGRS galaxies were added to the sample as optical hosts of 
NVSS doubles. These are listed at the end of Table 1, along with an unusual 
diffuse source, tentatively identified with TGN307Z092, which is discussed 
in Section 5.2. These nine extra galaxies bring the total number of candidate 
NVSS/2dFGRS matches to 912 (of these, 56\% are AGN and 35\% SF galaxies). 

\subsection{Optically--bright galaxies} 
We paid special attention to the brightest 2dFGRS galaxies ($b_{\rm J} =
14.0-17.0$\,mag).  For these bright galaxies, the radio--optical offset 
alone is not always a reliable indicator of whether the galaxy is associated 
with an NVSS radio source.  There are three reasons for this: 
(i) it can be difficult to measure an accurate optical position for 
these galaxies from Schmidt plates since the central regions are often 
over--exposed, (ii) we found that the 2dFGRS fibre positions used for bright 
galaxies were sometimes offset from what appeared to be the optical 
nucleus, and (iii) much of the radio emission in these galaxies arises 
from star formation--related processes in a resolved disk (see Section 5 
for more details), so that the centroid of the radio emission can be 
offset by several arcsec from the galaxy nucleus. 
We therefore inspected radio overlays on DSS images of all galaxies 
brighter than mag 17.0 and accepted those with 10.0---15.0 arcsec offsets 
as correct IDs if the radio emission appeared to be roughly centred on the 
optical galaxy. When this was done, it is noted in the final column of 
Table 1.

\subsection{The final sample } 
Our final sample of accepted NVSS/2dFGRS IDs comprises all sources with 
radio--optical offsets of 10.0\,arcsec or less, along with nine additional 
double galaxies described above and 47 additional galaxies with radio--optical 
offsets of 10.0--15.0 (28 of them radio sources associated with 
optically--bright galaxies, and 19 extended radio sources).  This gives a 
final sample of 757 `accepted' NVSS/2dFGRS IDs which we use in further 
analysis.  

\begin{figure}
\centering 
\vspace*{2cm}

{\it ** Fig. 5 available as a separate file ** } 

\vspace*{2cm}
 
\caption{Overlays of NVSS radio contours on DSS optical images of some of 
the extended radio sources identified with 2dFGRS galaxies. Each frame 
shows a region 6.4\,arcmin across.  Contours are at 1,2,4,8,16,32,...
\,mJy\,beam$^{-1}$,   
except for (h) TGS\,398Z220 and (i) TGS\,190Z105, where the lowest contour 
is at 4\,mJy\,beam$^{-1}$.  For TGS\,190Z105, two dark lines mark the position of the faint galaxy identified as the radio source. }
\end{figure}

\setcounter{table}{1}
\begin{table}
\caption{Summary of spectral classifications for the galaxies listed 
in Table 1}
\begin{tabular}{lrrr}
\hline
           & \multicolumn{1}{c}{All} &  \multicolumn{1}{c}{Accepted} & 
 \multicolumn{1}{c}{Rejected} \\
           &  \multicolumn{1}{c}{candidates} &  \multicolumn{1}{c}{as ID} & 
 \multicolumn{1}{c}{as ID} \\
\hline
 AGN       & 514        &  441     &   73  \\ 
 SF        & 319        &  272     &   47  \\
 Stars     &  18        &    0     &   18  \\
 Low S/N spectra&  61        &   44     &   17  \\
 Total     & 912        &  757     &  155  \\
\hline
\end{tabular}
\end{table}

Table 2 summarises the spectral properties of the final sample. 
The 2dFGRS spectra are generally of excellent 
quality, allowing both the redshift and spectral class to be determined 
accurately.  Of the 912 targets analysed here only 30 had spectra of 
quality class 1 or 2, i.e. were too noisy for a reliable redshift measurement. 
Eighteen targets turned out to be misclassified Galactic stars 
(i.e. had absorption--line spectra with $z < 0.001$), and two of the AGN 
were quasars (at redshifts $z=1.5$ and 3.0). 

We can use the data presented in Fig. 1 to place some quantitative limits 
on the reliability and completeness of the final sample. 
We consider three zones in radio--optical offset $\Delta$: (a) offsets below 
10\,arcsec, where all matches (other than Galactic stars) are accepted as 
correct IDs, (b) offsets of 10--15\,arcsec, where we accept only a subset  
of matches (those associated with bright optical galaxies or extended 
radio sources, where an eye inspection suggests that the match is a correct 
one), and (c) offsets greater than 15\,arcsec, which are rejected as 
correct IDs. 

In the first zone ($\Delta<10$\,arcsec) there are 712 matches, and we 
include 702 of these in the final sample. However, integrating under the 
line in Fig. 1 implies that $\sim75$ of these are likely to be chance 
coincidences. 
In the second zone ($\Delta=10-15$\,arcsec) there are 191 matches, of 
which 47 are included in the final sample.  We expect 106 chance 
coincidences, suggesting that the final sample should contain $\sim85$
objects in this zone rather than the 47 which are actually included. 
Thus our selection criteria have probably excluded some genuine IDs 
in this zone. 
In the third zone ($\Delta=15-30$\,arcsec) there are 590 matches, 
all of which we reject (we do however, include eight NVSS extended 
double radio sources, which were selected separately as described in 
Section 4.2).  Integrating under the line in Fig. 1 implies that we expect 
$\sim575$ chance coincidences in this zone, suggesting that there are few 
or no `missing' sample members in this zone. 

In summary, we estimate that our final sample of 757 radio--source IDs 
includes $\sim75$ objects which are chance coincidences of an 2dFGRs galaxy 
and a background radio source, and is missing $\sim40$ genuine 
radio-emitting galaxies which should have been included.  
Thus the sample is currently about 95\% complete and 90\% reliable 
(and the overall sample size is within 5\% of the correct value).  
In principle, both the compleness and the reliability could be raised 
to almost 100\% by measuring more accurate radio positions for the 
weaker (S$<5$\,mJy) radio sources in Table 1, and this will be done 
in the future.

\subsection{Radio stars?} 
As noted above, our final sample of radio--source IDs rejected all 
the NVSS--2dFGRS matches which had spectra characteristic of Galactic 
stars (with redshifts $z<0.001$).   As can be seen from Table 2, 
18 of the 155 rejected IDs ($12\pm3$\%) are Galactic stars.  
This is roughly twice the number expected for objects drawn at 
random from the 2dFGRS data set (6\% of the 2dFGRS objects observed 
up to March 2001 were Galactic stars; Colless et al.\ 2001), posing 
the question of whether we have detected any Galactic radio stars.  

At first glance, this seems unlikely.  Searches for radio stars in the 
NVSS and FIRST radio surveys (Condon et al.\ 1997; Helfand et al.\ 1999) 
find a very low detection rate at high Galactic latitudes (both studies 
find less than one radio--detected star per 200 deg$^2$ for V $< 10$\,mag, 
and Helfand et al.\ note that the detection probability drops steeply at 
fainter magnitudes).

Furthermore, the 2dFGRS stars are not randomly-selected stars but stars 
which have been misclassified as galaxies. Thus some of them are likely to  
be the chance superposition of a galaxy and a foreground star.  At least 
one object in Table 1 (TGN\,156Z046) certainly falls into this 
class -- it was originally classified as a star with 
$z=0.0005$, but examination of the spectrum showed that although the 
blue end of the spectrum was dominated by light from a foreground star, 
a higher-redshift system at $z=0.0391$ with emission lines of H$\alpha$,
[NII] and [SII] and Mg and NaD absorption lines could be seen clearly 
at the red end. 

Nevertheless, the superposition of galaxies and foreground stars is 
unlikely to provide a complete explanation for a larger--than--expected 
number of NVSS sources matched with Galactic stars 
because it is hard to understand why radio-emitting galaxies should be more 
likely than other 2dFGRS galaxies to be obscured by foreground stars.  
It will be interesting to see whether the `radio--star' excess persists 
as the 2dFGRS dataset grows. If so, it is possible that there may be a 
rare and so--far unrecognised class of radio sources associated with 
faint (b$_{\rm J} > 16$\,mag) Galactic stars.

\subsection{Other identification criteria} 
Machalski \& Condon (1999; MC99) took a different approach to 
cross--identifying NVSS radio sources with optical galaxies, and calculated 
a probability of association based on the radio--optical offset and the 
quoted errors in radio and optical positions.  They considered all NVSS 
radio sources within 30\,arcsec of an LCRS galaxy, and 
calculated a normalized offset 
\begin{equation}
\rho = [(\Delta_\alpha/\sigma_\alpha)^2 + (\Delta_\delta/\sigma_\delta)^2]\,
^{0.5} 
\end{equation}

\noindent
where $\Delta_\alpha$ and $\Delta_\delta$ are the differences between 
radio and optical positions and $\sigma_\alpha$ and $\sigma_\delta$ are 
the combined errors in the quoted radio and optical positions. They 
accepted a candidate identification as a true ID if $\rho < 2.5$.  
This method has the advantage that it allows a larger search radius for 
objects with larger position errors,  but the disadvantage is that the 
probability of finding a spurious optical ID is larger for faint sources 
because of the larger error box.  

We looked at the effect of applying the MC99 identification criterion to our
own sample. Of the 695 unresolved sources in Table 1, 543 are identified as IDs
by both criteria (i.e. MC99 and our own procedure as described above). 41
sources are rejected by both criteria, 38 are accepted by us but rejected by
MC99, and 73 are accepted by MC99 but rejected by us.  Thus the MC99 criterion
produces about 6\% more IDs than our method for the same data set. This is
probably not surprising --- we know that our method excludes a few 
genuine IDs with large radio--optical separation in order to produce a 
sample which is as uncontaminated by chance coincidences as possible.  

\subsection{K-corrections} 
The galaxies in our sample have redshifts as high as $z=0.3$ to 0.4, 
so when calculating galaxy luminosities we need to apply optical, radio and 
infrared K-corrections to take proper account of the effects of redshift 
on both the observed flux and the width of the passband. 

In the optical, we follow Folkes et al.\ (1999) and adopt 
the B--magnitude K-corrections tabulated by Pence (1976), using Pence's 
E/S0 values for our Aa and Ae galaxies, and the Sbc values for our 
SF galaxies. We also convert Pence's k(B) values to k($b_{\rm J}$) 
using the relation 
\begin{equation}
{\rm k}(b_{\rm J})={\rm k(B)}-0.28\,[{\rm k(B)}-{\rm k(V)}].  
\end{equation}
In the radio, we assume a mean spectral index of $\alpha = -0.7$ (where 
S$ \propto \nu^{\alpha}$) and apply the usual K-correction of the form 
\begin{equation}
{\rm K}(z)=(1+z)^{-(1+\alpha)}
\end{equation}
at redshift $z$. 

In the far--infrared, the situation is more complex because of the wide range 
in IRAS `spectral index' (i.e. dust temperature) observed in these 
galaxies (see Section 6).  Because the FIR K-correction can be either 
positive or
negative depending on what assumptions are made about the spectral energy 
distribution, and because many of our galaxies are weak IRAS sources with
detections in only one or two bands, we chose to apply no K-correction 
when calculating FIR luminosities. Since most of the FIR--detected galaxies 
we will study are at low redshift ($z<0.15$), the K-correction has little 
or no effect in any case. 

\section{Radio structures of resolved galaxies}
About 25\% of the radio sources associated with 2dFGRS galaxies are spatially
resolved by NVSS, allowing us to measure their projected linear size. In a few
cases we remeasured the angular size ourselves (usually because the NVSS source
catalogue split a single source into several components), otherwise we used the
NVSS catalogue value. Fig. 5 shows some of the extended NVSS sources 
identified with AGN and star--forming galaxies. 

Fig. 6 plots radio power against projected linear size for the 182 extended
NVSS radio sources with good--quality 2dFGRs spectra (i.e. with Q$\ge$3 in 
Table 1).  As expected, most star--forming galaxies are associated with radio
sources less than about 60\,kpc in diameter, i.e. no larger than a galaxy disk.
The few star--forming galaxies whose radio extent is larger than this appear to
be members of pairs or close groups in which more than one galaxy contributes
to the radio emission. In contrast, many of the radio sources with AGN spectra
are several hundred kpc in extent, consistent with classical (core plus jet)
radio galaxies. 

\begin{figure}

\vspace*{10cm}

\includegraphics{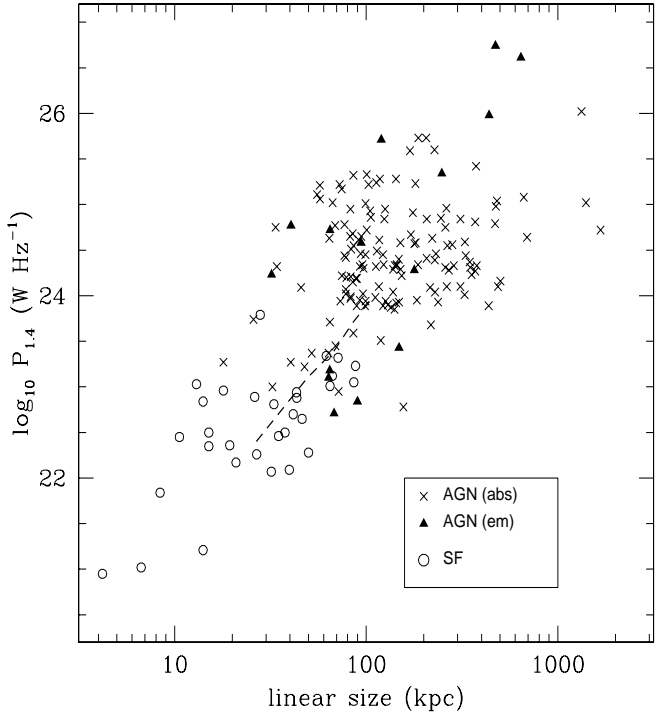}

\caption{Relationship between radio power and projected linear size for
spatially--resolved 2dFGRS/NVSS radio sources.  The dashed line shows the 
locus of a source with an observed angular size of 20\,arcsec and flux density
2.5\,mJy over the redshift range $z=0.05-0.3$.  Most of the upper limits 
for unresolved sources fall to the left of this line.  Sources 
larger than about 100\,kpc will be resolved throughout the 2dFGRS sample 
volume. } 
\end{figure} 

\subsection{Giant radio galaxies} 
Our sample includes three giant radio galaxies with projected linear sizes 
greater than 1\,Mpc, which are shown in Fig. 7.  Two are newly--discovered 
GRGs --- TGS223Z232 ($z=0.2095$) and TGS190Z081 ($z=0.2318$).  The third,
TGS241Z299, corresponds to the radio galaxy MRC\,B0312--271 ($z=0.2186$) 
which has already been identified as a GRG by Kapahi et al.\ (1998).  

About 50--60 GRGs are currently known (Ishwara--Chandra \& Saikia 1999; 
Schoenmakers et al.\ 2000) --- they are believed to represent the last 
stages of radio galaxy evolution, and to be unusually old or long--lived 
radio sources.  The three GRGs in our sample represent just under 1\% 
of the 441 AGN accepted as radio--source IDs, so they are clearly rare. 
It would be interesting to determine whether there is anything different 
about their environment (e.g. low density, lack of recent interactions with
companions) which has allowed these radio sources to grow undisturbed to 
such a large size. 

\begin{figure}
\centering 
\vspace*{2cm}

{\it ** Fig. 7 available as a separate file ** } 
\vspace*{2cm}

\caption{Radio contours overlaid on DSS optical images of the three giant 
radio galaxies (GRGs) detected in our sample: (a) TGS\,223Z232 (contours 1,2,3,4,5,10,20,50 mJy\,beam$^{-1}$), (b) TGS\,190Z081 
(contours 1,2,4,8,16,32,64,128 mJy\,beam$^{-1}$), (c) TGS241Z299 
(contours 1,2,5,10,20,50 mJy\,beam$^{-1}$). Each frame is 12.8\,arcmin 
across.  }
\end{figure}

\subsection{Unusual radio sources }
Fig. 8 shows two unusual radio sources discovered in our data set.  
The diffuse radio source shown in Fig. 8(a) is resolved into four separate
components in the NVSS catalogue, and one of these was originally picked up as
a possible match for the 2dFGRS galaxy TGN307Z090.  Inspection of the radio
contours showed that this is a large ($\sim6.5$\,\,arcmin diameter) source with
very low radio surface brightness.  

The source is clearly real, since it was independently detected at 
Green Bank (also at 1.4\,GHz) by White \& Becker (1992). 
The single--dish flux density of 189\,mJy suggests that NVSS may have 
missed some flux.  At present, nothing is known about the radio 
spectral index.   
While the identification is uncertain at this stage, the radio centroid 
is closer to the 2dFGRS galaxy TGN307Z092 than to its companion TGN307Z090, 
so we tentatively identify the source with TGN307Z092. At the redshift of 
TGN307Z092 ($z=0.0465$), the source has a projected linear size of 310\,kpc
and a 1.4\,GHz radio power of $1.3\times10^{24}$\,W Hz$^{-1}$.  Its radio
surface brightness is, however, unusually low.  The nature of the source
remains uncertain, though it may be a relic radio galaxy whose 
central engine has turned off (e.g. Komissarov \& Gubanov 1994). 

\begin{figure}
\centering 
\vspace*{2cm}

{\it ** Fig. 8 available as a separate file ** } 

\vspace*{2cm}

\caption{Radio contours overlaid on DSS optical images of two unusual radio 
sources in our sample.  Each frame is 8.6\,arcmin across.  (a) Overlay of 
NVSS radio contours on a DSS image of TGN\,307Z092 and its fainter companion 
TGN\,307Z090 (contour levels 1,2,3,4,6,8,10,12 mJy\,beam$^{-1}$), 
(b) NVSS contours overlaid on an image of TGS119Z122 (PKS\,2225--253; 
contour levels 1,2,4,6,8 mJy\,beam$^{-1}$).   }
\end{figure} 

Fig. 8(b) shows the radio emission associated with the 16th magnitude
star--forming galaxy TGS\,119Z122 at $z=0.0090$.  The NVSS flux density at
1.4\,GHz is 11.7\,mJy.  However, the same object is listed in the Parkes
catalogue as PKS\,2225--253, with flux densities of 230 and 130\,mJy at 
2.7 and 5.0\,GHz respectively.  The Parkes observations were made by 
Wall, Wright \& Bolton (1976), and the observing dates are given by them 
as 1974.0 and 1974.8 for the 2.7\,GHz and 5.0\,GHz observations respectively. 
Thus the source was detected at two different frequencies and on two 
different occasions separated by several months, and appears to have been 
real.  The optical identification with a 16th magnitude galaxy was first 
made by Savage \& Wall (1976). 

The radio luminosity of PKS\,2225--253 has declined dramatically 
since the 1974 Parkes observations.  One possibility, since TGS\,119Z122 
is actively forming stars, is that the bulk of the Parkes emission came 
from a radio--loud supernova like SN\,1986J or SN\,1998Z (e.g. Weiler et al.\
1998), which has since faded.  However, the early Parkes observations 
imply a 5 GHz radio power around $4\times10^{22}$ W\,Hz$^{-1}$, which is a 
factor of three higher than the peak luminosity of any known radio 
supernova including the gamma--ray burst object SN1998bw 
(Kulkarni et al.\ 1998).  A transient Galactic radio source seems
unlikely at $b=-58^\circ$, and the nature of the catalogued Parkes source
remains unclear. 

\section{IRAS detections and star--forming galaxies}
\subsection{IRAS detections of 2dFGRS galaxies} 
In the well--known correlation between far--infrared (FIR) and radio continuum 
emission in star--forming galaxies (e.g. de Jong et al.\ 1985, Helou et 
al.\ 1985), the FIR/radio ratio S$_{60\mu{\rm m}}$/S$_{1.49}$ has a mean
value of $\sim105$ (Condon \& Broderick 1988).  By a happy coincidence, this 
is very close to the ratio between the 280\,mJy limit of the IRAS Faint Source
Catalogue (FSC: Moshir et al.\ 1990) at 60\,$\mu$m and the 2.5\,mJy detection
limit of the NVSS at 1.4\,GHz.  As a result, we expect most star--forming
galaxies detected as radio sources by NVSS to be detected as 60\,$\mu$m sources
in the IRAS FSC, and vice versa. 

In practice, as noted earlier in Section 3.2, about 10\% of the 2dFGRS 
survey region was either unobserved by IRAS or had only single-scan coverage.  
Thus the absence of an individual galaxy from the IRAS catalogue does not necessarily mean that it has a 60\,$\mu$m flux below the FSC limit.  

Table 3 summarises the spectral properties of the 183 accepted radio--source 
IDs from Table 1 which were also detected at 60\,$\mu$m by IRAS.  Most galaxies 
detected at 60\,$\mu$m (83\%) were also detected by IRAS at 100$\mu$m.   

\begin{table}
\centering
\caption{Summary of spectral classifications for radio--source IDs which were 
also detected by IRAS } 

\begin{tabular}{lr}
\hline
  \multicolumn{1}{c}{2dF spectral class} &  \multicolumn{1}{c}{Galaxies } \\ 
\hline
 SF                    & 161  \\
 Ae                    &  16  \\ 
 Aae                   &   5  \\ 
 Low S/N               &   1  \\
 Total                 & 183  \\
\hline 
\end{tabular}
\end{table}

All the 2dF spectra of galaxies detected as IRAS sources show optical emission 
lines, in agreement with earlier studies (e.g. Allen et al.\ 1985, Lawrence et
al.\ 1986).  The great majority (89\%) are classified as SF, with the remainder
(11\%) being emission--line AGN with Seyfert--like spectra. 

Extrapolating from our current sample suggests that the full 2dFGRS database, 
when complete, will contain spectra of $\sim1000$ IRAS galaxies.  While this 
is smaller than targeted redshift surveys such as the IRAS PSCz (Saunders et 
al.\ 2000), which has more than 15,000 redshifts, it reaches to lower 
IRAS flux densities (and a higher median redshift) than most earlier surveys, 
as can be seen from Fig. 9.  
Since the general properties of IRAS galaxies are already well-explored by 
earlier studies, we focus here on the radio properties of star--forming 
galaxies in the 2dFGRS. 

\begin{figure}
\centering

\vspace*{4.8cm}

\vspace*{4.5cm}
\includegraphics{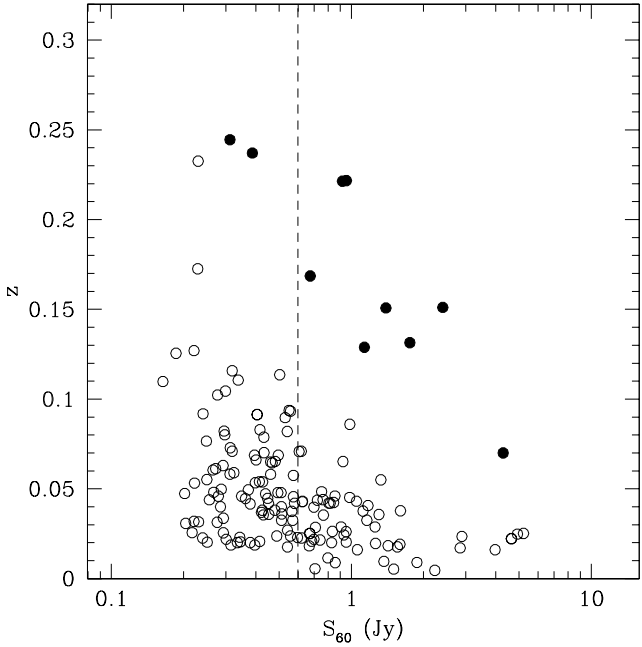}

\caption{Plot of IRAS 60\,$\mu$m flux density versus redshift for IRAS--detected 
2dFGRS radio sources.  ULIRGs with L$_{\rm FIR}>10^{12}$L$_\odot$ are shown 
by filled circles, and other galaxies by open circles.  The vertical line 
at 0.6\,Jy corresponds to the flux limit of the PSCz survey 
(Saunders et al.\ 2000).  } 
\end{figure} 

\subsection{The FIR--Radio Correlation} 
We calculated the FIR luminosity for each galaxy detected by IRAS, using 
the FIR flux defined by Helou et al.\ (1985): 
\begin{equation}
{\rm S}_{\rm FIR} = 1.26\times10^{-14} \times [2.58\,{\rm S}_{60}+{\rm S}_{100}], 
\end{equation}
where 
${\rm S}_{60}$ and ${\rm S}_{100}$ are the 60 and 100\,$\mu$m flux densities 
in Jy (1\,Jy=10$^{-26}$ W Hz$^{-1}$ m$^{-2}$) and S$_{\rm FIR}$ is in W m$^{-2}$.
  
Fig. 10 plots FIR luminosity versus 1.4\,GHz radio power for the 
IRAS--detected galaxies in our sample.  Most galaxies fall close to the
FIR--radio correlation for normal galaxies derived by Devereux \& Eales (1989),
\begin{equation}
\log_{10}{\rm P}_{1.49} = 1.28\,\log_{10}{\rm L}_{\rm FIR}+8.87, 
\end{equation}
but the scatter increases strongly for the most luminous galaxies. 

As can be seen from Table 4, the fraction of galaxies with AGN--like optical 
spectra increases with FIR luminosity and so it seems likely that the 
increased scatter in the FIR--radio correlation results from an increasingly 
diverse mix of pure star--forming galaxies and AGN or composite objects at  
higher FIR luminosities.  
\begin{figure}
\centering

\vspace*{5.2cm}


\vspace*{5cm}
\includegraphics{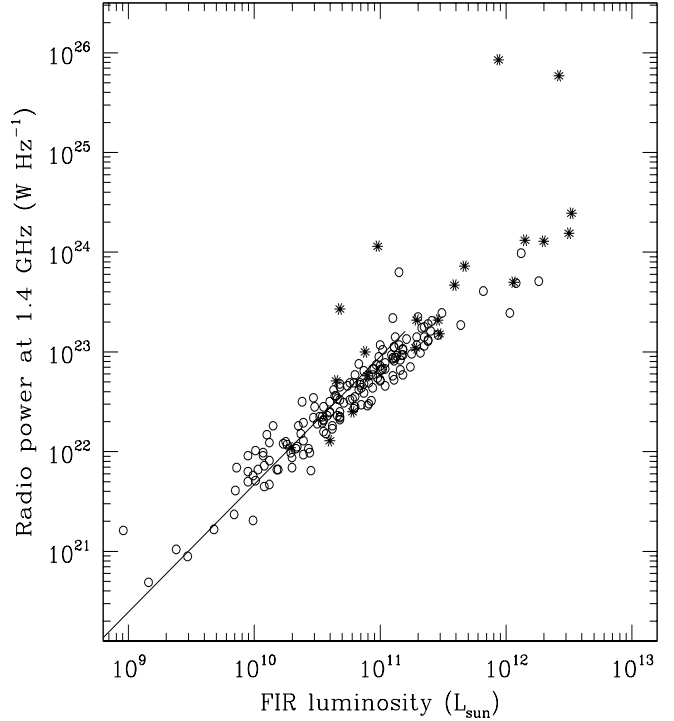}

\caption{Relation between FIR and radio continuum luminosity for radio--source
IDS from Table 1.  Galaxies with SF spectra are shown as open circles and those
with AGN spectra as stars.  The solid line shows the FIR--radio correlation
for normal galaxies derived by Devereux \& Eales (1989). Note that because of the requirement that they lie above the 2.8\,mJy NVSS detection limit, most of our 2dFGRS star--forming galaxies have star--formation rates which are significantly higher than the majority of `normal' galaxies. } 
\end{figure} 

\begin{table}
\centering
\caption{Summary of spectral properties for three bins in FIR luminosity. 
Hot IRAS galaxies are those with $\log_{10}$(S$_{60}$/S$_{100}$) $>-0.3$ 
(de Grijp et al.\ 1985) }
\begin{tabular}{crrrr}
\hline
  \multicolumn{1}{c}{$\log_{10}$(L$_{\rm FIR}$)} & \multicolumn{2}{c}{Fraction with }
 & \multicolumn{2}{c}{Fraction with } \\ 
  \multicolumn{1}{c}{(L$_{\odot}$)} &  \multicolumn{2}{c}{AGN spectra } 
& \multicolumn{2}{c}{hot IRAS colours } \\ 
\hline
 $>$11.5    & 60\% & (9/15)  & 92\% & (11/12)  \\
 10.5--11.5 &  9\% & (11/118) & 46\% & (45/99) \\
 $<$10.5    &  2\% & (1/49)   & 27\% & (11/41) \\
\hline 
\end{tabular}
\end{table}

\subsection{IRAS colours and 2dF spectral types} 
As noted by de Grijp et al.\ (1985), galaxies with active nuclei tend to have 
hotter IRAS colours (as measured by the flux ratio S$_{60}$/S$_{100}$) 
than `normal' star--forming galaxies, and galaxies with Seyfert 
nuclei generally have S$_{60}$/S$_{100}>0.5$. 

Fig. 11 shows a FIR `colour--magnitude diagram' for galaxies in our 
sample.  The IRAS flux densities for weaker sources have typical 
errors of 10--15\%, so the IRAS colours can have uncertainties of 30\% or more. 
However, there is still a general increase in the fraction of galaxies with 
`hot' IRAS colours at higher FIR luminsity (see also the summary in Table 4), 
consistent with an increasing contribution from active nuclei at higher FIR 
luminosity. 

\begin{figure}
\centering

\vspace*{4.6cm}


\vspace*{4.6cm}
\includegraphics{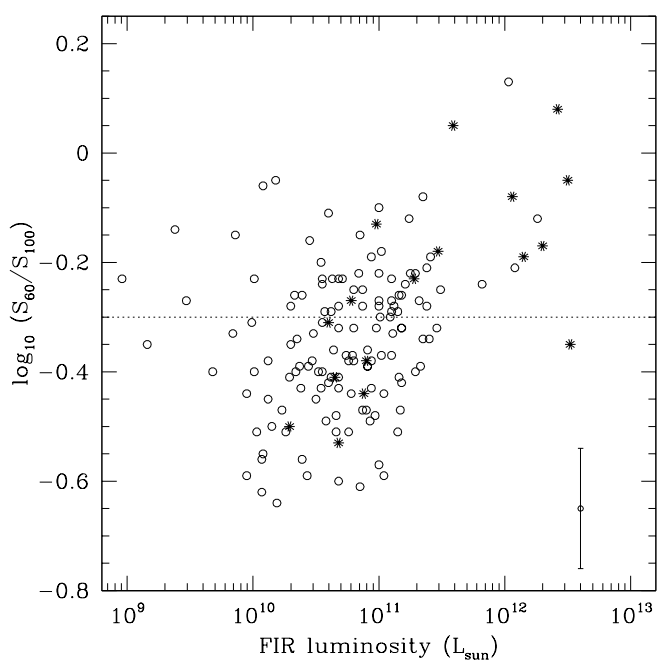}

\caption{Plot of IRAS `colour' ($\log_{10}$(S$_{60}$/S$_{100}$) versus FIR luminosity 
for galaxies in our sample.  `Warm' galaxies are defined by de Grijp et al. 
(1985) as those with S$_{60} > 0.5$\,S$_{100}$, and they find that most 
Seyfert galaxies have `warm' IRAS colours. The boundary between warm and 
cool galaxies as defined by de Grijp et al.\ is shown by the dashed horizontal 
line, and a typical error bar for the weakest IRAS sources 
(S$_{60}\simeq0.3$\,Jy) is shown in the lower right--hand corner. 
As in Fig. 10, open circles and stars denote galaxies classified spectroscopically 
as SF and AGN respectively. }
\end{figure} 

\begin{table*}
\centering
\begin{minipage}{170mm} 
\caption{Ultra--luminous IRAS Galaxies (ULIRGs) in the sample } 
\begin{tabular}{@{}llccccrcl@{}}
\hline
2dFGRS name & IRAS name & Spectral &  $z$ & Abs. & $\log_{10}$P$_{1.4}$      & $\log_{10}$ & $\log_{10}$L$_{\rm FIR}$ & Notes \\
           &            &  class   &          & mag. & (W\,Hz$^{-1}$) & S$_{60}$/S$_{100}$ & (L$_\odot$)  &  \\
\hline
 TGS206Z015  &  IRAS  00335--2732  & SF  & 0.0700 & --20.71 & 23.39  &    0.13 & 12.03  & S92; Megamaser \\
 TGS209Z156  &  IRAS  00482--2720  & SF  & 0.1289 & --21.04 & 23.69  &  --0.21 & 12.08  & K98, V99, C99\\
 TGS238Z241  &  IRAS  03000--2719  & Ae  & 0.2214 & --22.52 & 24.39  &  --0.35 & 12.52  & C96 \\
 TGN152Z171  &  IRAS F09521--0400  & Ae  & 0.2371 & --22.03 & 24.12  &  --0.19 & 12.15  & New ULIRG\\
 TGN314Z018  &  IRAS  11598--0112  & Ae  & 0.1511 & --22.01 & 24.19  &  --0.05 & 12.50  & S92; ROSAT \\
 TGN131Z280  &  IRAS  12532--0322  & Ae  & 0.1686 & --22.18 & 23.70  &  --0.08 & 12.06  & New ULIRG \\
 TGN137Z043  &  IRAS  13270--0331  & Ae  & 0.2217 & --22.16 & 25.77  &    0.08 & 12.42  & C95 \\
 TGN206Z237  &  IRAS  14121--0126  & Ae  & 0.1508 & --21.35 & 24.11  &  --0.17 & 12.30  & C95 \\
 TGS178Z172  &  IRAS  22206--2715  & SF  & 0.1314 & --21.98 & 23.71  &  --0.12 & 12.26  & C96 \\
 TGS180Z060  &  IRAS F22301--2822  & SF  & 0.2445 & --21.92 & 23.99  & $>$--0.47& 12.12 & C96 \\
\hline 
\end{tabular} 

References:  C95: Clowes et al.\ (1995); C96 Clements et al.\ (1996);  C99 Clements et al.\ (1999); 
K98 Kim \& Saunders (1998); \\S92: Strauss et al.\ (1992); V99 Veilleux et al.\ (1999)
\end{minipage}
\end{table*}

\subsection{Ultraluminous IRAS galaxies} 
Ten of the IRAS galaxies in our sample (listed in Table 5) have FIR 
luminosities above 10$^{12}$ L$_\odot$, and can be considered 
`ultraluminous IRAS galaxies' (ULIRGS; e.g. Sanders \& Mirabel 1996).  
Most of these galaxies are already known as ULIRGs from other surveys, 
but there are two newly--discovered ULIRGs, TGN\,152Z171 and TGN\,131Z280. 
Fig. 12 shows their spectra. 

All the galaxies in Table 5 have unresolved radio sources in the NVSS, though 
because of the relatively large distances of these galaxies this only sets 
fairly unrestrictive limits (typically 100--200\,kpc) on the linear size of 
the associated radio emission. 

\begin{figure}
\centering

\vspace*{4cm}


\vspace*{4cm}
\includegraphics{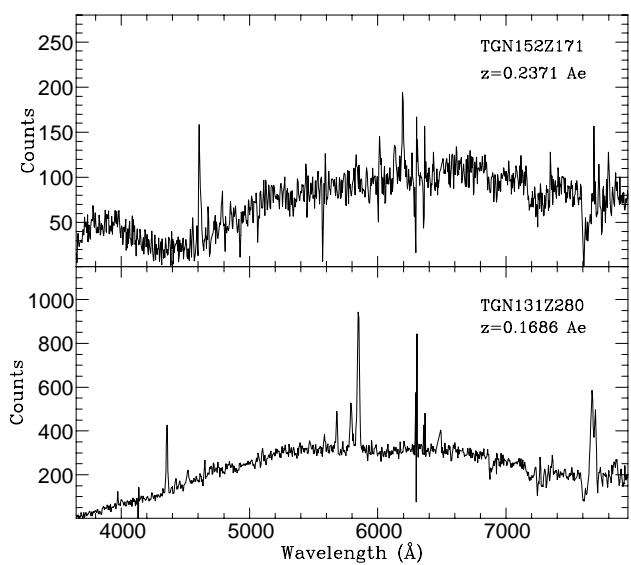}

\caption{2dFGRS spectra of the two newly-dscovered ULIRGs in our sample, 
TGN152Z171 ($B_{\rm J}=18.85$) and TGN131Z280 ($B_{\rm J}=17.93$). 
Both spectra show residuals due to incomplete subtraction of the 
atmospheric airglow line at 6300\AA. }
\end{figure} 

\section{ROSAT detections}
Bauer et al.\ (2000) have cross--matched the NVSS source catalogue with the 
ROSAT Bright Source Catalogue (RSBC) of X--ray sources 
(Voges et al.\ 1999).  They showed that the relatively low surface density 
of NVSS sources allows reliable identification of the radio counterparts of 
RSBC sources despite uncertainties of 10\,arcsec or more in the X--ray 
positions.  The more accurate radio positions then allow a reliable optical 
identification to be attempted.  Bauer et al.\ showed that the RSBC/NVSS 
radio sources were dominated by AGN with an average redshift of $\langle z\rangle\simeq0.1$. 

\begin{figure}
\centering

\vspace*{5.0cm}


\vspace*{5.0cm}
\includegraphics{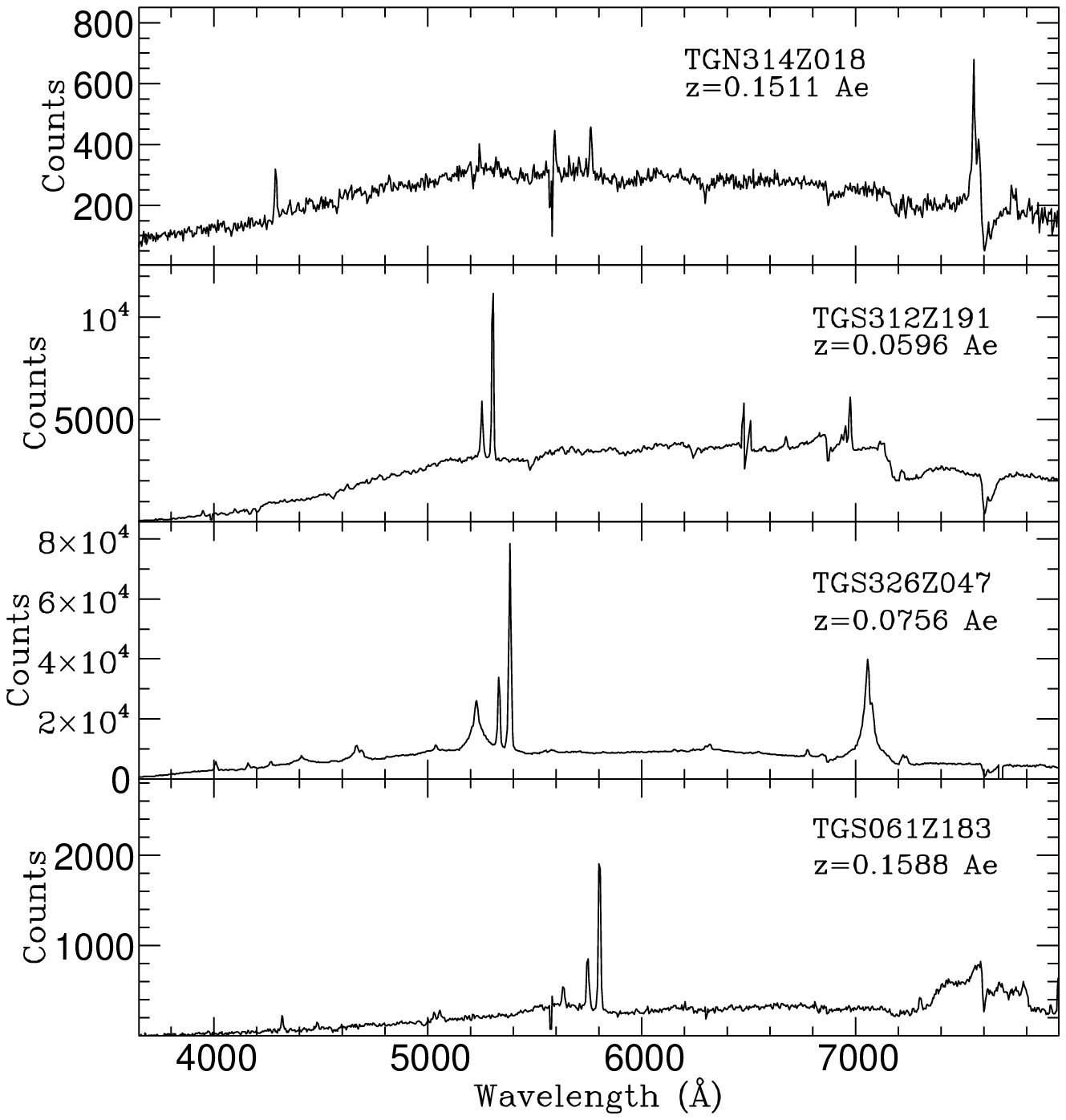}

\caption{2dFGRS spectra of four NVSS galaxies detected as X--ray sources 
in the ROSAT All--Sky Survey.  All show optical emission lines with line 
ratios characteristic of active galaxies, but as noted in the text they 
are otherwise a diverse class of objects.  }
\end{figure} 
 
We cross--matched the galaxies in Table 1 with the ROSAT All--Sky Survey (RASS) 
catalogue, taking a simple cutoff of 30\,arcsec as our matching radius rather 
than the more complex criteria prescribed by Bauer et al.\ (2000).  This gave 
the six matches listed in Table 6, four of which are also in the Bauer et 
al.\ list (the other two fall below the 0.1 ct s$^{-1}$ X--ray limit imposed 
by Bauer et al.\ for their sample). 

All six galaxies in Table 6 have strong emission lines and spectra classed 
as Ae, but in other respects they are a diverse population, including three 
powerful, double--lobed radio galaxies, two Seyfert 1 galaxies with less powerful radio emission, and one ultra--luminous IRAS galaxy (ULIRG, see 
Table 5 and Fig.\ 13). 

Bauer et al.\ identified 1512 extragalactic RSBC/NVSS sources with Galactic 
latitude $|b|>15^\circ$ in the 10.3\,sr NVSS survey area.  This corresponds 
to a surface density of about 0.05 objects per square degree, or about four 
times higher than the corresponding surface density derived from the 
RSBC/NVSS/2dFGRS objects in Table 6 (0.012 objects per square degree).  

Where are the missing RSBC/NVSS objects?  We expect most of them to fall 
within the 2dFGRS volume, since Bauer et al.\ note that their objects are 
mostly local with a mean redshift of about 0.1.  Closer examination of the 
Bauer et al.\ objects in the 2dFGRS area suggests that about half the missing 
objects are bright, nearby galaxies which were excluded from the 2dFGRS.  
Most of the remainder appear to be Seyfert galaxies excluded because of 
their compact, stellar appearance on sky survey plates, along with a few 
higher--redshift QSOs. 

\begin{table*}
\begin{minipage}{170mm} 
\caption{Galaxies detected as X--ray sources in the ROSAT All--Sky 
Survey (RASS) } 
\begin{tabular}{@{}llccrcll@{}}
\hline
2dFGRS name & ROSAT name &         Spectral &  $z$ & Abs. & 
$\log_{10}$P$_{1.4}$ & RSBC count & Notes \\
            &            &         class    &      & mag. & (W\,Hz$^{-1}$) & rate (s$^{-1}$)&  \\
\hline
TGS312Z191 & 1RXS\,J023513.9$-$293616 & Ae & 0.0596 &  --22.43 &  24.58 & 0.36$\pm$0.03 & Radio galaxy MRC\,0234--287 \\
TGN314Z018 & 1RXS\,J120226.9$-$012908 & Ae & 0.1511 &  --22.01 &  24.19 & 0.14$\pm$0.03 & ULIRG \\
TGN401Z254 & 1RXS\,J133253.5$+$020047 & Ae & 0.2162 &  --22.14 &  26.74 & 0.25$\pm$0.03 & Radio galaxy PKS\,1330$+$02  \\
TGS326Z047 & 1RXS\,J214636.3$-$305132 & Ae & 0.0756 &  --21.48 &  22.84 & 0.84$\pm$0.07 & Seyfert 1 \\
TGS407Z205 & 1RXS\,J215809.2$-$312341 & Ae & 0.0933 &  --23.01 &  23.17 & 0.08$\pm$0.02 & Seyfert 1 \\
TGS061Z183 & 1RXS\,J220924.2$-$245326 & Ae & 0.1588 &  --21.51 &  25.98 & 0.05$\pm$0.02 & Radio galaxy PKS\,2206--251  \\
\hline 
\end{tabular}
\end{minipage}
\end{table*}

\section{The local radio luminosity function (RLF) at 1.4\,GHz}
\subsection{Calculation of the local RLF}
We now calculate the radio luminosity function (RLF) for 2dFGRS--NVSS 
galaxies with $z \le 0.3$, both 
for the sample as a whole and for the AGN and SF subclasses. We assigned
spectral types to the 19 unclassified galaxies with measured redshifts as
follows: galaxies with $b_{\rm J}$ fainter than magnitude 17.0 and radio 
power above 10$^{23}$ W\,Hz$^{-1}$ were assigned to the Aa 
(absorption-line AGN) class, otherwise they were are assumed to be 
star--forming (SF).  Using these criteria, we classify 18 of the low S/N galaxies as Aa and only one as SF. 

To calculate the local RLF, we use the $1/{\rm V}_{\rm max}$ method 
(Schmidt 1968), as discussed by Condon (1989).  V$_{\rm max}$ is calculated 
from the joint optical and radio limits of the sample, i.e. a radio cutoff 
of 2.8\,mJy and optical cutoffs at $b_{\rm J}=14.0$\,mag (bright) and 
19.4\,mag (faint).  Table 7 lists the derived local RLF for the 
whole sample, and for the AGN and SF classes separately. At this stage, 
we make no corrections for incompleteness and the only normalisation 
is the effective area of 325\,deg$^2$ derived in Section 2.1.  

Fig. 14 shows the derived RLF, together with earlier values derived by 
Condon (1989) and Machalski \& Godlowski (2000).  An advantage of the 
2dF/NVSS sample is that all the data are drawn from a single radio survey 
and a set of homogeneous optical spectra from a single instrument. 
Most previous determinations of the local RLF used data from several 
radio surveys to span the equivalent range in radio power.  
The good overall agreement with earlier derivations confirms our earlier 
calculation that the incompleness in our final sample is small ($<10$\%). 

\begin{table*}
\centering
\begin{minipage}{140mm} 
\caption{The local radio luminosity function at 1.4\,GHz }
\begin{tabular}{@{}crcrcrc@{}}
\hline
   & \multicolumn{2}{c}{All galaxies } & \multicolumn{2}{c}{AGN} & 
   \multicolumn{2}{c}{Star--forming galaxies } \\ 
$\log_{10}$L$_{1.4}$ & N & \multicolumn{1}{c}{$\log_{10}\Phi$} & N & 
\multicolumn{1}{c}{$\log_{10}\Phi$} & N & \multicolumn{1}{c}{$\log_{10}\Phi$} \\ 
 (W Hz$^{-1}$) &   & (mag$^{-1}$ Mpc$^{-3}$)& & (mag$^{-1}$ Mpc$^{-3}$) & & (mag$^{-1}$  Mpc$^{-3}$) \\
\hline
   26.6    &  1  & $-7.96^{+0.30}_{-1.0}$ &  1 & $-7.96^{+0.30}_{-1.0}$ &  &  \\
   26.2    &  1  & $-8.01^{+0.30}_{-1.0}$ &  1 & $-8.01^{+0.30}_{-1.0}$ &  &  \\
   25.8    &  8  & $-6.86^{+0.13}_{-0.19}$ &  8 & $-6.86^{+0.13}_{-0.19}$ &  &  \\
   25.4    & 14  & $-6.57^{+0.10}_{-0.14}$ & 14 & $-6.57^{+0.10}_{-0.14}$ &  &  \\
   25.0    & 21  & $-6.38^{+0.09}_{-0.11}$ & 21 & $-6.38^{+0.09}_{-0.11}$ &  &  \\
   24.6    & 57  & $-5.87^{+0.05}_{-0.06}$ & 57 & $-5.87^{+0.05}_{-0.06}$ &  &  \\
   24.2    & 79  & $-5.72^{+0.05}_{-0.05}$ & 75 & $-5.74^{+0.05}_{-0.05}$ &  
  4 & $-6.98^{+0.18}_{-0.30}$ \\
   23.8    &110  & $-5.46^{+0.04}_{-0.04}$ &101 & $-5.49^{+0.04}_{-0.05}$ &  
  9  & $-6.64^{+0.12}_{-0.18}$ \\
   23.4    & 93  & $-5.14^{+0.04}_{-0.05}$ & 66 & $-5.32^{+0.05}_{-0.06}$ &  
 27  & $-5.62^{+0.08}_{-0.09}$ \\
   23.0    &106  & $-4.56^{+0.04}_{-0.04}$ & 44 & $-4.98^{+0.06}_{-0.07}$ &  
 62  & $-4.77^{+0.05}_{-0.06}$ \\
   22.6    & 80  & $-4.14^{+0.05}_{-0.05}$ & 22 & $-4.73^{+0.08}_{-0.10}$ &  
 58  & $-4.27^{+0.05}_{-0.06}$ \\
   22.2    & 58  & $-3.70^{+0.05}_{-0.06}$ &  8 & $-4.52^{+0.13}_{-0.19}$ &  
 50  & $-3.77^{+0.06}_{-0.07}$ \\
   21.8    & 28  & $-3.49^{+0.08}_{-0.09}$ &  2 & $-4.69^{+0.23}_{-0.53}$ &  
 26  & $-3.52^{+0.08}_{-0.09}$ \\
   21.4    &  4  & $-3.59^{+0.18}_{-0.30}$ &    &  &  4  & $-3.51^{+0.18}_{-0.30}$ \\
   21.0    &  1  & $-3.74^{+0.30}_{-1.00}$ &    &  &  1  & $-3.74^{+0.30}_{-1.0}$ \\
   20.6    &  1  & $-3.26^{+0.30}_{-1.00}$ &    &  &  1  & $-3.26^{+0.30}_{-1.0}$ \\ 
\hline
Total & 662 &  &  420 & &  242 & \\
$\langle$V/V$_{\rm max}\rangle$ &  
\multicolumn{2}{l}{0.528$\pm$0.011} 
 & \multicolumn{2}{l}{0.542$\pm$0.014 } 
 & \multicolumn{2}{l}{0.503$\pm$0.019 } \\
\hline
\end{tabular}
\end{minipage}
\end{table*}

\begin{figure}

\vspace*{5cm}


\vspace*{5cm}
\includegraphics{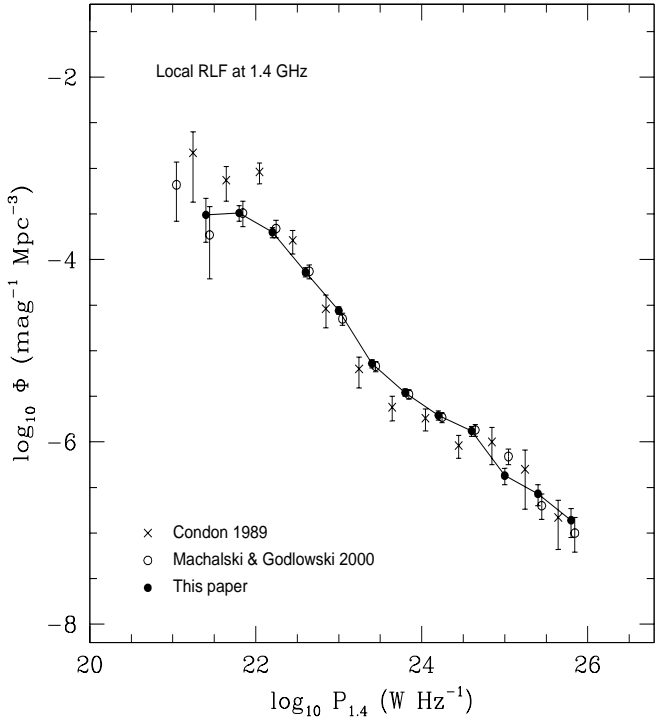}

\caption{Local RLF derived from the 662 galaxies in Table 1 which are 
accepted as correct IDs and have radio flux density S$_{\rm 1.4}\ge2.8$\,mJy, 
optical magnitude $14.0\le b_{\rm J}\le19.4$ and redshift $0.001<z<0.3$.  
Previous derivations by Condon (1989) and Machalski \& Godlowski (2000) are 
shown for comparison.  Between 10$^{22}$ and 10$^{24}$\,W\,Hz$^{-1}$, our 
values and those of Machalski \& Godlowski are sometimes so close that they 
are indistinguishable in the diagram. }
\end{figure} 

\begin{figure}

\vspace*{4.5cm}


\vspace*{4.5cm}
\includegraphics{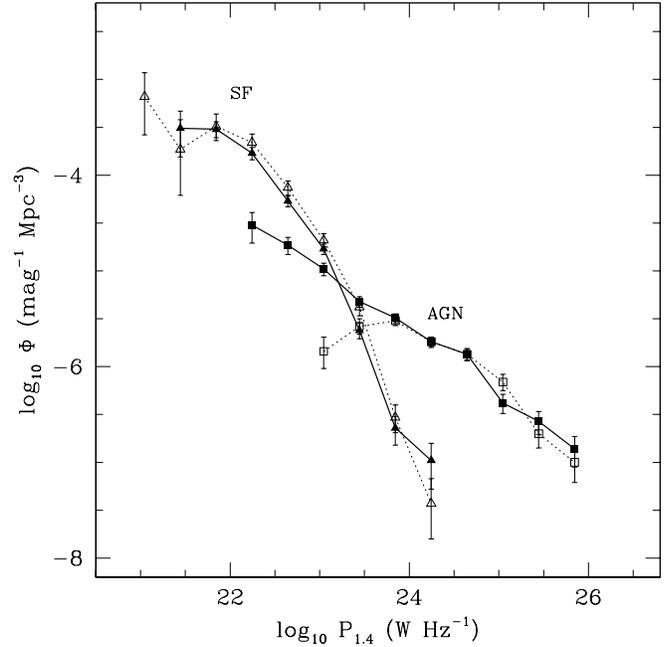}

\caption{Local RLF derived for AGN and SF galaxies separately. Filled points 
connected by solid lines show our own data for AGN (squares) and SF galaxies 
(triangles). The local RLFs for AGN and SF galaxies derived by Machalski \& Godlowski (2000) are plotted using open symbols connected by dotted lines. }

\end{figure} 

\subsection{Comparison with earlier work} 
Figures 14 and 15 shows a comparison between our values of the local 
RLF and those derived by Machalski \& Godlowski (2000) for a sample of 
1157 radio--identified galaxies from the LCRS.  This is the only other determination which uses a comparably--large data set of homogeneous 
optical spectra. 

As can be seen from Fig. 14, the overall agreement between our total 
(i.e. AGN plus SF) RLF and that of MG is extremely good.  This is remarkable 
given the differences between the two samples: 
\begin{itemize}
\item 
Our criteria for determining radio--source IDs differ from those used by 
Machalski \& Condon (1999; hereafter MC99), as noted in Section 4.6.
\item
All the 2dFGRS redshifts were determined spectroscopically, whereas most 
of the redshifts listed by MC99 were estimated from optical magnitudes 
(so that some individual values may have large errors). 
\end{itemize}
Such good agreement between two independent samples chosen in different ways  
suggests that the RLF is a very robust indicator of the overall density 
of radio sources in the local universe.  

As can be seen from Fig. 15, however, the agreement between our results 
and those of Machalski \& Godlowski (2000; hereafter MG) breaks down 
when we split our sample into AGN and SF galaxies. 
We find that the space density of AGN radio sources continues to rise 
as we go to radio powers as low as 10$^{22}$ W\,Hz$^{-1}$, with no sign 
of a turnover (see Fig. 15).  In contrast, MG find a decreasing density 
of radio AGN below 10$^{24}$ W\,Hz$^{-1}$, which is reflected in the 
divergence of the two AGN LFs in Fig. 15.  This turnover in the MG LF 
for AGN has also been discussed by Brown et al.\ (2001), who ascribe it to 
incompleteness in the AGN data used by MG.  The good agreement between 
the faint end of our local RLF for AGN and that derived by Sadler 
et al.\ (1989) for nearby galaxies can be seen in Figure 18, and 
strongly supports the view that our values are correct and that the MG 
sample is incomplete for low--luminosity AGN.  


Thus, although we and MG agree on the overall density of radio sources 
in the local universe, we disagree on the relative fraction of AGN and 
SF galaxies below 10$^{24}$ W\,Hz$^{-1}$.  There are two possible 
explanations for this, a {\it selection difference}\ (i.e. the different 
criteria for inclusion in the two samples select roughly the same 
{\it number}\ of galaxies, but do not necessarily select the same 
{\it kind} of galaxies), and a {\it classification difference}. 
Our AGN/SF classification is based on optical spectra, whereas MG used 
the classifications from MC99, which take into account several factors 
including the radio--optical flux ratio, angular extent of the radio 
emission, and IRAS data where available. 

There are 92 galaxies in common between our set of radio--source IDs 
in Table 1 and the LCRS data set used by MC99.  Of these, we agree on the 
classification of 69 (40 AGN, 29 SF), i.e. 75\% of the galaxies in common. 
Of the 25\% of galaxies where there is disagreement, most are classified 
by us as AGN and by MC99 as SF (i.e. for a data set classified by both 
groups, we find more AGN, and fewer SF galaxies, than MC99). 

A detailed comparison of the 2dFGRS and LCRS data sets is outside the 
scope of this paper, but it seems likely that both selection differences 
and misclassification of low--power AGN as SF galaxies contribute to the 
incompleteness of the MG AGN data at low radio powers.  The identification 
criteria used by Machalski \& Condon (1999) may exclude some genuine 
radio IDs with flux densities above 5\,mJy and radio--optical position 
offsets of less than 10\,arcsec (see Section 4.6), and the radio--optical 
flux ratio which they use in their classification may misclassify some 
low--power AGN as SF galaxies (especially since the radio sources in 
low--power AGN are rarely spatially resolved by NVSS, and so cannot be 
recognised by their radio morphology in the same way as many powerful 
radio AGN).

\subsection{Star--forming galaxies and the local star--formation density } 

\begin{figure}

\vspace*{5cm}


\vspace*{5cm}
\includegraphics{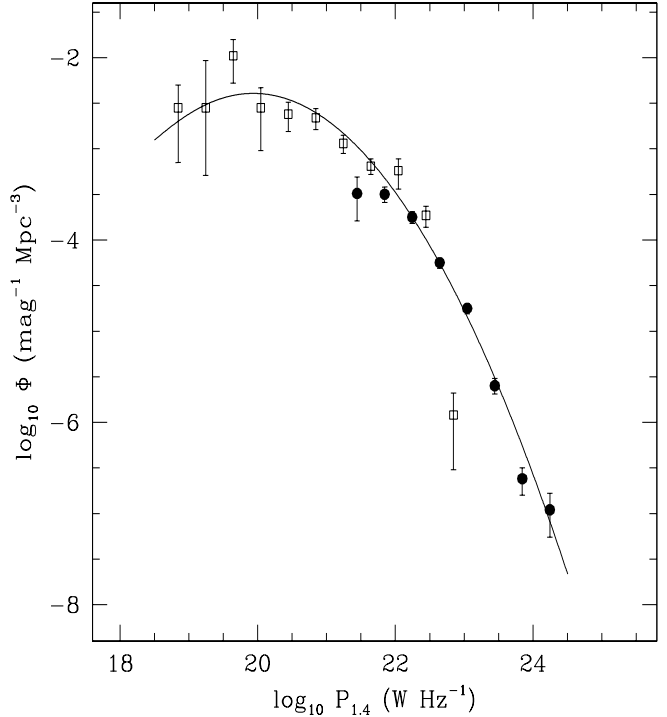}

\caption{Local RLF for star--forming galaxies, combining data from our 
sample (filled circles) with values from Condon (1989) for nearby 
star--forming galaxies from the Revised Shapley Ames Catalogue (open squares). 
The NVSS/2dFGRS values of $\Phi$ have been increased by 5\% to take into 
account the 95\% redshift completeness of the 2dFGRS.  The solid line shows 
the best--fitting analytic function as described in the text.  }

\end{figure} 

As noted earlier, the 2dFGRS excluded most bright, nearby galaxies with 
$b_{\rm J} < 14$\,mag.   To extend our sample to lower 
radio luminosity, we therefore combined our measured RLF 
with the local RLF derived by Condon (1989) for galaxies in the Revised 
Shapley Ames Catalogue (RSA; Sandage \& Tammann 1981).  In doing this, we 
also increased the values of $\Phi$ 
listed in Table 7 by 5\% to correct for the $\sim5$\% spectroscopic 
incompleteness of the 2dFGRS (i.e. the spectra in Table 1 with Q$\leq2$ 
for which no reliable redshift could be measured).  As can be seen from 
Fig. 16, our results agree well with the Condon RSA values in the small 
region of overlap.  We then fitted an analytic funtion of the type described 
by Saunders et al.\ (1990): 
\begin{equation}
\Phi ({\rm L}) = {\rm C}\  \left[{\frac{\rm L}{\rm L_\star}}\right]^{1-\alpha} 
{\rm exp} \lgroup - \frac{1}{2} 
\left[ \frac{\log_{10}(1+{\rm L}/{\rm L_\star})}{\sigma}\right]^2 \rgroup 
\end{equation} 
\noindent
to the combined data, and Table 8 summarizes the results.  

\begin{table}
\centering
\caption{Parametric fits to the local RLFs for AGN and SF galaxies, using the 
Saunders et al.\ (1990) fitting function as described in the text }
\begin{tabular}{cccl}
\hline
  \multicolumn{1}{c}{Parameter } & \multicolumn{1}{c}{SF galaxies }
 & \multicolumn{1}{c}{AGN } & Units \\ 
\hline
 ${\rm log}_{10}$L$_\star$  &  19.55$\pm$0.03    &  24.59$\pm$0.03  & W Hz$^{-1}$ \\
 $\alpha$       &  0.840$\pm$0.020  &  1.58$\pm$0.02   &  \\
 $\sigma$       &  0.940$\pm$0.004  &  1.00$\pm$0.13   & \\
 $\log_{10}$C          &  --2.41$\pm$0.04 &  --5.89$\pm$0.02 & mag$^{-1}$ Mpc$^{-3}$ \\   
 $\chi^2$       &   1.86         &   0.91       &  \\       
\hline 
&&& \\
\end{tabular}
\end{table}

We can now use the RLF for star--forming galaxies to estimate the 
local star--formation density (i.e. the zero--point of the Madau diagram; 
Madau 1996).  Following Cram et al.\ (1998) and Haarsma et al.\ (2000), 
we assume a Salpeter--like initial mass function 
\begin{equation}
\Psi({\rm M}) \propto {\rm M}^{-2.35} 
\end{equation}
over the range 0.1 to 100\,M$_\odot$, 
and convert from a radio luminosity to a star--formation rate (SFR) 
via the relation 
\begin{equation}
{\rm SFR}\ ({\rm M}_\odot\,{\rm yr}^{-1}) = \frac{{\rm L}_{1.4} 
\ ({\rm W Hz}^{-1})}{8.85\times10^{20}} 
\end{equation} 

\noindent
(Sullivan et al.\ 2001).  The local star--formation density at any given SFR 
is then 
\begin{equation} 
{\large \rho}_{\rm SF} = {\rm SFR}\,({\rm L}_{1.4})\ \times\ \Phi\,({\rm L}_{1.4}) 
\end{equation} 

\noindent
where $\Phi$ is the local RLF from Tables 7 and 8, multiplied by 1.05 to 
correct for incompleteness as noted above.  Fig. 17 shows the 
results --- our data imply that the greatest contribution to the local 
star--formation density comes from galaxies with star--formation rates 
around 10 M$_\odot$ yr$^{-1}$.  

As can be seen from Fig. 17, our radio--derived values for the 
local star--formation density are in excellent agreement with the 
values derived from H$\alpha$ by Gallego et al.\ (1995; hereafter 
G95) for galaxies with star--formation rates up to 
20--30 M$_\odot$ yr$^{-1}$.  

For galaxies with the highest star--formation rates ($>$30\,M$_\odot$ yr$^{-1}$),  
however, we find a significantly higher density than G95.  The reasons for 
this are not completely clear --- our SF galaxies with high derived SFRs 
appear to be genuine star--forming galaxies which follow the FIR--radio 
correlation (see Fig. 10). Where measurements of diagnostic emission--line 
ratios have been carried out on the 2dF spectra, these also 
confirm the SF classification (Jackson \& Londish 2000).  

Our sample volume for galaxies with a high SFR is larger than that surveyed 
by G95.  Their survey covered 471 deg$^{2}$ to a depth of $z\leq0.045$ 
(beyond which the H$\alpha$/[NII] lines were shifted out of their passband), 
i.e. a maximum volume of about 9$\times10^5$ Mpc$^{3}$.  The equivalent 
volume for the SF galaxies in our current (325 deg$^{2}$) sample is set by 
the redshift at which the observed radio flux density falls below our 2.8\,mJy 
cutoff.  For a galaxy with a star--formation rate of $\sim100$\,M$_\odot$ 
yr$^{-1}$, this redshift is $z\simeq0.084$, giving a volume of 
$4.5\times10^6$\,Mpc$^3$, or five times the volume surveyed by G95.  

Because of this increase in sample volume for stronger radio 
sources, the 2dFGRS/NVSS SF sample is dominated by galaxies with high star--formation rates (well over half the SF galaxies we detect have 
derived star--formation rates above 30 M$_\odot$ yr$^{-1}$, compared 
to only 5\% of the G95 galaxies) so we would expect to have better 
statistics than G95 for galaxies with high SFRs, assuming that the radio luminosity continues to scale linearly with star--formation rate. 

It is possible that at high star--formation rates the 
H$\alpha$ emission line is increasingly obscured by dust, so that optical 
surveys underestimate the number of galaxies with very high star--formation 
rates.  Deep VLA studies of galaxy clusters at $z\sim0.4$ (Smail et 
al.\ 1999) and local galaxies with `post--starburst' optical spectra 
(Miller \& Owen 2001) suggest that some galaxies may have star--forming 
regions which are largely hidden by dust.  Follow--up observations of the 
galaxies in Table 1 for which the radio data imply high star--formation 
rates would therefore be valuable. 

Integrating under the curve in Fig. 17 gives a local star--formation 
density of $0.022\pm0.004$ M$_\odot$ yr$^{-1}$ Mpc$^{-3}$, which is slightly 
higher than the value of $0.013^{+0.007}_{-0.005}$ derived by G95 from 
H$\alpha$ data. The difference arises mainly because our sample contains 
more galaxies with high star--formation rates ($>$30\,M$_\odot$ yr$^{-1}$).  
For galaxies with star--formation rates up to 50\,M$_\odot$ yr$^{-1}$, 
we derive a local density of $0.017\pm0.004$ M$_\odot$ yr$^{-1}$ Mpc$^{-3}$, 
in excellent agreement with the G95 value. 

\begin{figure}

\vspace*{4cm}


\vspace*{5cm}
\includegraphics{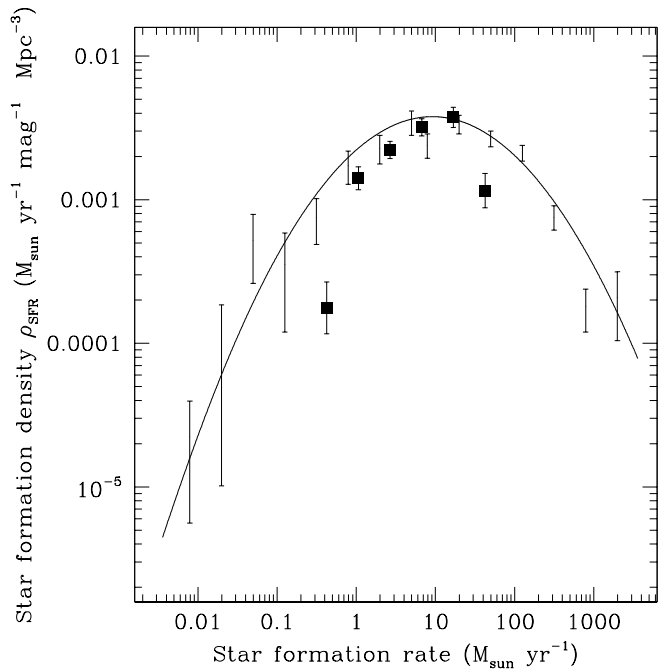}

\caption{Local star--formation density (in M$_\odot$ yr$^{-1}$ mag$^{-1}$ Mpc$^{-3}$) 
for galaxies with star--formation 
rates between 0.01 and 1000 M$_\odot$ yr$^{-1}$.  The solid line is derived 
from the fit to the local RLF for star--forming galaxies (Table 8) after 
correcting for spectroscopic incompleteness.  
Error bars from the individual data points used to derive the fit are also 
shown.  Filled squares show the values of local star--formation density 
derived from H$\alpha$ measurements by Gallego et al.\ (1995).  }
\end{figure}

\subsection{Active galaxies and radio galaxies} 

\begin{figure}

\vspace*{5cm}


\vspace*{5cm}
\includegraphics{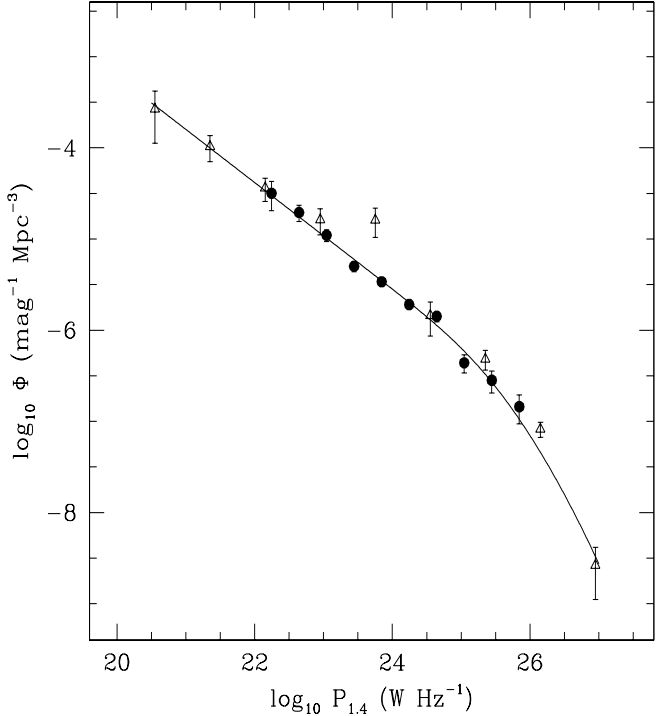}

\caption{Local RLF for active galaxies, combining data from our 
sample (filled circles) with values from Sadler et al.\  (1989) for 
nearby E/S0 galaxies (open triangles).  The Sadler et al.\ values have 
been coverted to 1.4\,GHz assuming a spectral index of $\alpha=-0.7$. 
As in Fig. 16, the NVSS/2dFGRS values of $\Phi$ have been increased by 
5\% to take into account the 95\% redshift completeness of the 2dFGRS, 
and the solid line shows the best--fitting analytic function (see Table 8). }
\end{figure} 

As in Section 8.3, we combined the NVSS/2dFGRS sample with published data 
for bright, nearby galaxies to extend our results to lower radio power.  
Fig. 18 shows the results --- the NVSS/2dFGRS data points agree well 
with the RLF for nearby (B $< 14.0$\,mag) elliptical and S0 
galaxies from Sadler et al.\ (1989).  Once again, we fitted an analytic 
function as described in Section 8.3, and the results are given in Table 8.  
However, it is remarkable that the space density of radio--emitting AGN 
is also well fitted by a single power--law of the form 
\begin{equation}
\Phi({\rm P_{1.4}}) \propto {\rm P_{1.4}}^{-0.62\pm0.03}
\end{equation}
over almost 
five decades in luminosity from $10^{20.5}$ to $10^{25}$ W Hz$^{-1}$, 
before turning down above $10^{25}$ W Hz$^{-1}$.  
As pointed out by Sadler et al.\ (1989), the AGN RLF must also turn down 
below $10^{20}$ W Hz$^{-1}$ in order not to exceed the space density of 
luminous galaxies. 

\subsection{Black holes in radio AGN} 
Franceschini, Vercellone \& Fabian (1998) examine the relation between 
galaxy lumninosity, black hole mass and radio power in nearby active 
galaxies, and conclude that the radio power of an AGN is both a good 
tracer of supermassive black holes and an estimator of their mass 
(though Laor (2000) notes that the correlation between radio power and 
black hole mass shows considerable scatter). 

Following the precepts of Franceschini et al., we can estimate the 
local mass density of black holes from our AGN RLF with the following 
conversion factors: 
\begin{equation}
\log_{10}{\rm M}_{\rm BH}\ ({\rm M}_\odot) = 0.376\ \log_{10}{\rm P}_{1.4}\ 
({\rm W\,Hz}^{-1}) + 0.173 
\end{equation}
 
\noindent 
and 
\begin{equation}
\log_{10}\Phi_{\rm BH}\ =\ \log_{10}\Phi_{1.4} + 0.425 
\end{equation}

\begin{figure}

\vspace*{5cm}


\vspace*{5cm}
\includegraphics{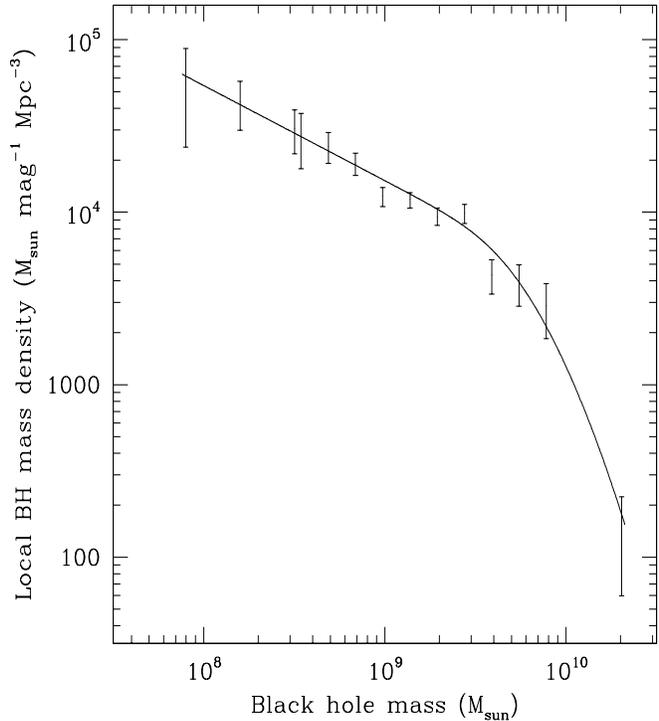}

\caption{Local mass density (in M$_\odot$\,mag$^{-1}$\,Mpc$^{-3}$) of 
black holes, derived from the local 
radio luminosity function for AGN (Tables 7 and 8) together with the 
mean relation between radio power and black hole mass from Franceschini 
et al.\  (1998). As in Fig. 17, error bars are shown for the individual 
points used to derive the fit in Table 8.  }
\end{figure} 

This yields the mass density distribution shown in Fig. 19.  In contrast 
to the star--formation density plot shown in Fig. 17, the total mass  
density of black holes does not converge but continues to increase down 
to the lowest values (a few times $10^7$ M$_\odot$) so far probed by radio 
surveys.  Integrating over the values in Fig. 19 gives 
a total mass density of massive black holes (M$_{\rm BH} > 7.6\times10^7$\,
M$_\odot$) in galactic nuclei of 
\begin{equation}
\rho_{\rm BH}\ =\ 1.8^{+0.4}_{-0.6}\times10^5\ {\rm M}_\odot\ {\rm Mpc}^{-3} 
\end{equation}
  
\noindent
which is within the range $1.4-2.2\times 10^5$ derived by Chokshi \& Turner 
(1992) from the optical luminosity function of QSOs.  We note that the value derived here is actually a lower limit, since the derived black--hole 
mass density is still increasing at the lowest values of M$_{\rm BH}$ we can measure.  Thus a comparison of black--hole mass densities for radio galaxies 
in the local universe and high--redshift QSOs suggests that local 
radio--emitting AGN are the direct descendants of most or all of the 
high--$z$\ QSOs.  

\subsection{Redshift evolution of the radio luminosity function} 
Although the 2dFGRS probes to redshifts of $z\sim 0.3$ to 0.4 where we 
might expect to see cosmic evolution of the radio--source population, 
only the most luminous objects in our sample can be seen to these distances, 
and hence we can only test for evolutionary effects over a narrow range 
in luminosity.  There are already hints that we are seeing 
evolution in the most powerful AGN in our sample -- Table 9 shows the mean values of V/V$_{\rm max}$ for AGN and star--forming galaxies split into bins 
in radio luminosity. 
For AGN with $\log_{10}$P$_{1.4} > 10^{23}$ W Hz$^{-1}$, 
$\langle$V/V$_{\rm max}\rangle$ is significantly higher than the expected 
value of 0.50, implying that the space density of these objects is higher 
at higher redshift.  

Because the number of objects which we can use to probe evolutionary effects 
is small, we defer discussion of the RLF evolution to a later paper in 
this series which will analyse the full set of 2dFGRS data.

\begin{table}
\caption{Values of $\langle$V/V$_{\rm max}\rangle$ from RLF calculations split into 
bins in radio power} 
\begin{tabular}{crcrc}
\hline
 $\log_{10}$P$_{1.4}$  & \multicolumn{2}{c}{----- AGN -----} &  
\multicolumn{2}{c}{----- SF -----} \\
 (W Hz$^{-1}$)  & n & $\langle$V/V$_{\rm max}\rangle$  & n & 
$\langle$V/V$_{\rm max}\rangle$ \\
\hline
 22--23    &   46 &  0.50$\pm$0.05 &  141 & 0.48$\pm$0.03   \\
 23--24    &  195 &  0.55$\pm$0.02 &   65 & 0.54$\pm$0.04   \\ 
 24--25    &  145 &  0.55$\pm$0.02 &    4 & 0.53$\pm$0.12   \\
\hline
\end{tabular}
\end{table}

\section{Discussion and conclusions}
\subsection{Main results} 
We have shown how combining data from large radio continuum and optical 
redshift surveys allows us to derive accurate local radio luminosity 
functions (RLFs) for AGN and star--forming (SF) galaxies.  Both AGN and 
star--forming galaxies are significant contributors to the local RLF 
below $10^{24}$ W\,Hz$^{-1}$ (at higher radio powers, almost all the 
sources are AGN), so good-quality optical spectra are needed to classify 
the radio sources correctly. 

This paper establishes an accurate local benchmark for future studies 
of the cosmic evolution of both AGN and star--forming galaxies at 
higher redshift. The full data set of $\sim4000$ radio--source spectra 
which will become available when the 2dFGRS is completed should be 
large enough to measure the evolution of radio galaxies to $z=0.35$ and 
the most luminous star--forming galaxies to $z=0.2$. 

\subsection{2dFGRS radio--source populations needing further investiagation } 
We showed in Section 6 that there may be a substantial local population of 
radio--luminous star--forming galaxies (with implied star--formation rates 
of 50\,M$_\odot$\,yr$^{-1}$ or more) which are not seen in H$\alpha$ 
emission--line surveys.  Determining whether these ``high SFR'' galaxies 
are dust--enshrouded starbursts or misclassified AGN is important if we are 
to have an accurate census of star formation in the local universe (if they 
are indeed starbursts, the ``high SFR galaxies'' contribute about 20\% of 
the local star--formation density).  High--resolution radio imaging, together 
with infrared spectroscopy, should allow us to determine whether the 
radio emission seen by NVSS arises mainly from dusty star--forming regions.  

Optical spectra of the remaining NVSS/ROSAT sources in 
the 2dFGRS area (see Section 7) would be valuable in determining whether some 
radio AGN have been excluded from the 2dFGRS because they have a bright 
nucleus which leads to them being classified as stars rather than galaxies 
on sky survey plates.  We estimate that no more than five or six potential 
members of the current sample have been excluded in this way, but it would 
be useful to confirm this. 

\subsection{Prospects for deeper radio and optical observations in the 
2dFGRS area } 
Since the overlap between 2dFGRS galaxies and NVSS radio sources is 
relatively small --- about 5\% of NVSS radio sources in the 2dFGRS area
are associated with 2dFGRS galaxies, and 1--2\% of 2dFGRS galaxies are 
detected by NVSS ---  it is tempting to speculate on what could be 
achieved with deeper radio and optical observations in the 2dFGRS area.  

\begin{figure}

\vspace*{10cm}
\includegraphics{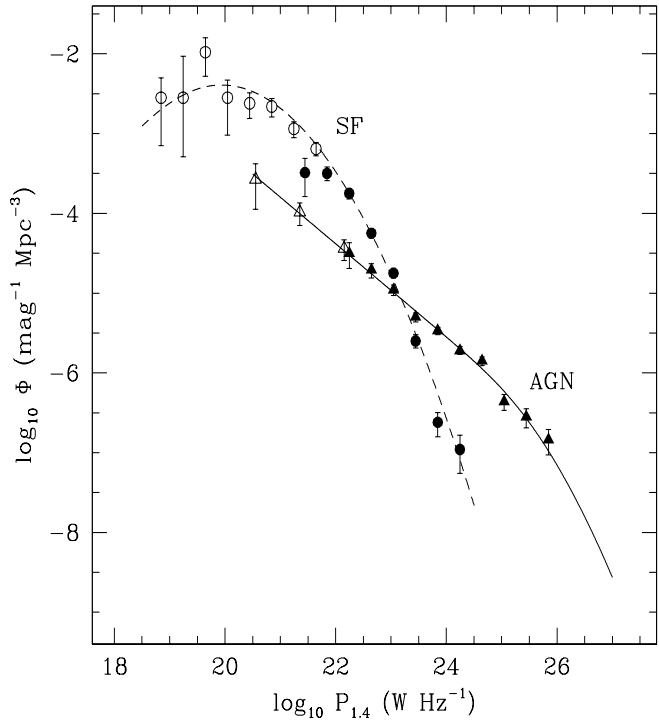}

\caption{Local radio luminosity functions for AGN and star--forming galaxies, 
combining 2dFGRS data with nearby RSA spirals (SF) from Condon (1989) and 
nearby E/S0 galaxies (AGN) from Sadler et al.\ (1989).  Note that both AGN and 
SF galaxies contribute significantly at all radio powers below $\sim10^{25}$ W Hz$^{-1}$ at 1.4\,GHz.} 

\end{figure}

Radio observations to sub--mJy sensitivities at 1.4\,GHz would increase 
the detection rate for galaxies in the 2dFGRS sample, particularly for 
star--forming galaxies since many of these lie close to the 2.8\,mJy 
NVSS detection limit (see Section 3.4).  For example, observations 
with a 3$\sigma$ 
detection limit of 0.4\,mJy at 1.4\,GHz could detect galaxies with 
a star--formation rate of $\sim17$\,M$_\odot$\,yr$^{-1}$ at $z=0.1$, 
which is significantly lower than the limit of $\sim120$\,M$_\odot$\,yr$^{-1}$ 
at the same redshift for galaxies near the NVSS detection limit of 2.8\,mJy. 

About 30\% of all NVSS radio sources have an optical counterpart visible 
on the digitised sky survey (i.e. brighter than $b_{\rm J}\sim23$\,mag). 
Deep 2dF spectroscopy should be possible to $b_{\rm J} = 21$\,mag 
(and even fainter for emission--line objects) with integration times of 
4--8 hours rather than the 40--60 minutes 
used by the 2dFGRS, and careful attention to sky--subtraction techniques 
(Cannon 2001).  This would allow spectroscopy of the host galaxies of 
powerful AGN (and hence studies of their cosmic evolution) to redshifts of $z\sim0.7$ rather than the $z\sim0.35$ limit of the 2dFGRS. 

\begin{figure}

\vspace*{10cm}
\includegraphics{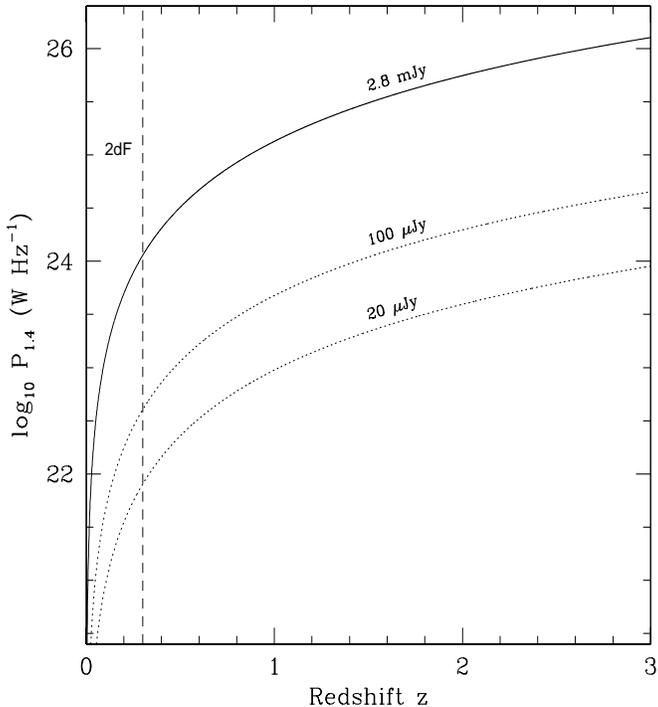}

\caption{Radio luminosity limits at high redshift for NVSS (2.8\,mJy limit)  
and deeper 1.4\,GHz surveys (100\,$\mu$Jy and 20\,$\mu$Jy limits).  
The vertical line at $z$=0.3 marks the effective redshift limit of the 2dFGRS. 
Note that at $z\sim1$ the deepest surveys so far carried out 
detect galaxies with radio luminosities above about $10^{23}$\,W\,Hz$^{-1}$, 
i.e. in the regime where we expect a mixture of AGN and star--forming
galaxies. } 
\end{figure}

\subsection{Implications for deep radio surveys to $\mu$Jy levels }
Fig. 20 shows the local radio luminosity functions (RLFs) for star--forming  
(SF) galaxies and AGN over the whole range in radio power for which data 
are currently available.  The two RLFs cross over (i.e. AGN and SF galaxies 
contribute equally to the local radio--source population) at P$_{1.4} \simeq 
10^{23.2}$ W\,Hz$^{-1}$.  Both AGN and SF galaxies contribute at least 
20\% of the radio--source population over the range $10^{22.5} - 10^{23.7}$\,W\,Hz$^{-1}$. 

Fig. 21 shows the radio luminosity range probed at different redshifts 
by 1.4\,GHz surveys with flux--density limits of 2.8\,mJy (NVSS), 
100\,$\mu$Jy (e.g. Gruppioni at al.\ 1997, Hopkins et al.\ 1998) and 
20\,$\mu$Jy (e.g. Haarsma 2000).  Extrapolating from the local RLF 
suggests that all $\mu$Jy--level radio surveys should find a mixture of AGN 
and star--forming galaxies at all redshifts, i.e. there is no observational 
regime in which we can simply assume that a faint radio source arises 
from a starburst. Deep radio continuum surveys in the Hubble Deep Field 
(Richards et al.\ 1998; Garrett et al.\ 2000) suggest that the $\mu$Jy 
radio--source population is composed of a mixture of 70--90\% 
spiral and irregular/merging galaxies and 10--30\% ellipticals. 

A mixture of star--forming galaxies and AGN is probably present even 
at flux densities below 1$\mu$Jy  --- Hopkins et al.\ (2000) remark that 
in their simulations of the faint radio--source population to a flux density limit of 0.1\,$\mu$Jy, the proportion of AGN is still significant.  If this 
is true, then optical/IR spectroscopy will play a key role in disentangling 
the faint radio--source population probed by future deep radio surveys, 
and partnerships between radio and optical telescopes in mapping out 
the distant universe will increase in importance in the years to come.  

\section{Acknowledgments}
The 2dF Galaxy Redshift Survey was made possible through the dedicated 
efforts of the staff of the Anglo--Australian Observatory, both in 
creating the 2dF instrument and in supporting it on the telescope. 
This research has made use of the NASA/IPAC Extragalactic Database 
(NED) which is operated by the Jet Propulsion Laboratory, Caltech,
under contract with NASA.  We thank Prof.\ Lawrence Cram for helpful 
conversations about the derivation of star--formation rates from radio 
data, and the referee, Dr I.\ Snellen, for several perceptive comments 
which improved the final version of this paper.


\begin{thebibliography}{99}  
\bibitem{} Allen, D.A., Roche, P.F., Norris, R.F. 1985, MNRAS, 213, 67P 
\bibitem{} Bauer, F.E., Condon, J.J., Thuan, T.X., Broderick, J.J., 
   2000, ApJS, 129, 547 
\bibitem{} Becker, R.H., White, R.L., Helfand, D.J. 1995, ApJ, 450, 559 
\bibitem{} Beichman, C.A., Neugebauer, G., Habing, H.J., Clegg, P.E., 
  Chester, T.J. 1988, IRAS Catalogs and Atlases, Version 2. Explanatory 
  Supplement
\bibitem{} Bock, D.C-J., Large, M.I., Sadler, E.M. 1999,  AJ, 117, 1593 
\bibitem{} Brown, M.J.I., Webster, R.L., Boyle, B.J. 2001, AJ, 121, 2381 
\bibitem{} Cannon, R., 2001, AAO Newsletter, 96, 13 
\bibitem{} Chokshi, A., Turner, E.L. 1992, MNRAS, 259, 421 
\bibitem{} Clements, D.L., Sutherland, W.J., Saunders, W., Efstathiou, G.P., 
  McMahon, R.G., Maddox, S., Lawrence, A., Rowan-Robinson, M. 1996, MNRAS, 
  279, 459  
\bibitem{} Clements, D.L., Saunders, W.J., McMahon, R.G. 1999, MNRAS, 302, 391 
\bibitem{} Clowes, R.G., Campusano, L.E., Leggett, S.K., Savage, A. 1995, 
  MNRAS 275, 819  
\bibitem{} Colless, M. 1999, Phil Trans R Soc Lon A, 357, 105 
\bibitem{} Colless, M. et al. 2001, MNRAS, in press (astro-ph/0106498)   
\bibitem{} Condon, J.J., Broderick, J.J. 1988, AJ, 96,30  
\bibitem{} Condon, J.J. 1989,  ApJ, 338, 13 
\bibitem{} Condon, J.J. 1992,  ARAA, 30, 575 
\bibitem{} Condon, J.J., Kaplan, D.L., Yin, Q.F. 1997, AAS, 191.1402   
\bibitem{} Condon, J.J., Cotton, W.D., Greisen, E.W., Yin, Q.F., Perley, R.A., 
   Taylor, G.B., Broderick, J.J. 1998,  AJ, 115, 1693 
\bibitem{} Cram, L.E. 1998, ApJ, 506, 85
\bibitem{} de Grijp, M.H.K., Miley, G.K., Lub, J., de Jong, T. 1985, 
  Nature, 314, 240  
\bibitem{} de Jong, T., Klein, U., Wielebinski, R., Wunderlich, E 1985, A\&A, 
     147, L6 
\bibitem{} Devereux, N.A., Eales, S.A. 1989, ApJ, 340, 708 
\bibitem{} Folkes, S., et al. 1999, MNRAS, 308, 459 
\bibitem{} Franceschini, A., Vercellone, S., Fabian, A.C. 1998, MNRAS, 297, 817
\bibitem{} Gallego, J., Zamorano, J., Aragon--Salamanca, A., Rego, M. 1995, 
      ApJ, 455, L1
\bibitem{} Garrett, M.A., de Bruyn, A.G., Giroletti, M., Baan, W.A., 
      Schilizzi, R.T. 2000, A\&A, 361, L41 
\bibitem{} Gruppioni, C., Zamorani, G., de Ruiter, H.R., Parma, P., Mignoli, 
     M., Lari, C. 1997, MNRAS, 286, 470 
\bibitem{} Haarsma, D.B., Partridge, R.B., Windhorst, R.A., Richards, E.A. 
  2000, ApJ, 544, 641
\bibitem{} Helfand, D.J., Schnee, S., Becker, R.H., White, R.L., McMahon, 
  R.G. 1999, AJ, 117, 1568 
\bibitem{} Helou, G., Soifer, B.T., Rowan--Robinson, M, 1985, ApJ, 298, L7 
\bibitem{} Hopkins, A.M., Mobasher, B., Cram, L., Rowan--Robinson, M. 1998, 
   MNRAS, 296, 839 
\bibitem{} Hopkins, A., Windhorst, R., Cram, L., Ekers, R. 2000, Experimental 
   Astronomy, 10, 419 
\bibitem{} Ishwara--Chandra, C.H., Saikia, D.J. 1999, MNRAS, 309, 100 
\bibitem{} Jackson, C.A., Londish, D.M. 2000, PASA, 17, 234 
\bibitem{} Jauncey, D.L. 1975, ARAA, 13, 23 
\bibitem{} Kapahi, V.K., Athreya, R.M., van Breughel, W., McCarthy, P.J., 
   Subrahmanya, C.R. 1998, ApJS, 118, 275 
\bibitem{} Kim, D.-C., Sanders, D.B. 1998, ApJS, 119, 41 
\bibitem{} Komissarov, S.S., Gubanov, A.G. 1994, A\&A, 285, 27 
\bibitem{} Kulkarni, S.R., Frail, D.A., Wieringa, M.H., Ekers, R.D., 
  Sadler, E.M., Wark, R.M. Higdon, J.L., Phinney, E.S., Bloom, J.S. 
  1998, Nature, 395, 663 
\bibitem{} Laor, A. 2000, ApJ, 543, L111 
\bibitem{} Lawrence, A., Walker, D., Rowan--Robinson, M., Leech, K.J., 
   Penston, M.V. 1986, MNRAS, 219, 687 
\bibitem{} Lewis, I.J. et al. 2001, in preparation 
\bibitem{} Longair, M.S. 1966, MNRAS, 133, 421 
\bibitem{} Machalski, J., Condon, J.J. 1999, ApJS, 123, 41 
\bibitem{} Machalski, J., Godlowski, W., 2000, A\&A, 360, 463 
\bibitem{} Madau, P., Ferguson, H.C., Dickinson, M.E., Giavalisco, M., 
   Steidel, C.C., Fruchter, A. 1996, MNRAS, 283, 1388
\bibitem{} Madgwick, D.\ et al.\ 2001, MNRAS, submitted (astro-ph/0107197)  
\bibitem{} Magliocchetti, M., Maddox, S.J., Lahav, O., Wall, J.V. 1998, 
  MNRAS, 300, 257
\bibitem{} Miller, N.A., Owen, F.N. 2001, ApJ, 554,L25  
\bibitem{} Moshir, M., et al.\ 1990, IRAS Faint Source Catalogue Version 2.0 
\bibitem{} Pence, W. 1976, ApJ, 203, 39 
\bibitem{} Rengelink, R.B., Tang, Y., de Bruyn, A.G., Miley, G.K., Bremer, 
   M.N., R\"ottgering, H.J.A., Bremer, M.A.R. 1997, A\&AS, 124, 259
\bibitem{} Richards, E.A., Kellermann, K.I., Fomalont, E.B., Windhorst, R.A., 
   Patridge, R.B. 1998, AJ, 116, 1039 
\bibitem{} Sadler, E.M., Jenkins, C.R., Kotanyi, C.G. 1989, MNRAS, 240, 591 
\bibitem{} Sadler, E.M., McIntyre, V.J., Jackson, C.A., Cannon, R.D. 1999, 
 PASA, 16, 247 (Paper I)
\bibitem{} Sandage, A., Tammann, G.A., 1981, {\it A Revised Shapley--Ames 
 Catalogue of Bright Galaxies}, Washington, Carnegie Inst. 
\bibitem{} Sanders, D.B., Mirabel, I.F. 1996, ARAA, 34, 749 
\bibitem{} Saunders, W., Rowan--Robinson, M., Lawrence, A., Efstathiou, G., 
 Kaiser, N., Ellis, R. S., Frenk, C. S., 1990,  MNRAS, 242, 318 
\bibitem{} Savage, A., Wall, J.V. 1976, AuJPA, 39, 39 
\bibitem{} Schoenmakers, A.P., Mack, K.-H., de Bruyn, A.G., R\"ottgering,  
 H.J.A., Klein, U., van der Laan, H. 2000, A\&AS, 146, 322 
\bibitem{} Schmidt, M. 1968, ApJ, 151, 393 
\bibitem{} Smail, I., Morrison, G., Gray, M.E., Owen, F.N., Ivison, R.J., 
 Kneib, J.-P. 1999, ApJ, 525, 609 
 Ellis, R. S.
\bibitem{} Strauss, M.A., Huchra, J.P., Davis, M., Yahil, A., Fisher, K.B., 
 Tonry, J. 1992, ApJS, 83, 29 
\bibitem{} Sullivan, M., et al.\ 2001, submitted to MNRAS 
\bibitem{} Veilleux, S., Kim, D.-C., Sanders, D.B. 1999, ApJ, 522, 113 
\bibitem{} Voges, W., Aschenbach, B., Boller, T., Br\"auninger, H., Briel, U., 
 Burkert, W., Dennerl, K., Englhauser, J., Gruber, R., Haberl, F., Hartner, G., 
 Hasinger, G., K\"urster, M., Pfeffermann, E., Pietsch, W., Predehl, P., 
 Rosso, C., Schmitt, J.H.M.M., Tr\"umper, J., Zimmermann, H.U. 1999, A\&A, 
 349, 389 
\bibitem{} Wall, J.V., Wright. A.E., Bolton, J.G. 1976, AuJPA, 39, 1
\bibitem{} Wall, J.V., Pearson, T.J., Longair, M.S. 1980, MNRAS, 193, 683 
\bibitem{} Weiler, K.W., Montes, M.J., Panagia, N., Sramek, R.A. 1998, ApJ, 
   500, 51 
\bibitem{} White, R.L., Becker, R.H. 1992, ApJS, 79, 331 
\bibitem{} York, D.G. et al. 2000, AJ, 120, 1579 
\end{thebibliography}
\end{document}